# New ternary ThCr$_2$Si$_2$-type iron-selenide superconducting materials: synthesis, properties and simulations.

A.L. Ivanovskii

*Institute of Solid State Chemistry, Ural Branch of the Russian Academy of Sciences, 620990, Ekaterinburg, Russia*

A B S T R A C T

Very recently (November, 2010, *PRB, 82, 180520R*), the first ThCr$_2$Si$_2$-type ternary superconductor K$_{0.8}$Fe$_2$Se$_2$ with enhanced $T_C$ ~ 31 K has been discovered. This finding has stimulated much activity in search for related materials and triggered intense studies of their properties. Indeed, very soon superconductivity ($T_C$ ~ 28-32 K) was also found in the series of related ternary systems (so-called 122 phases) such as Cs$_x$Fe$_{2-y}$Se$_2$, Rb$_x$Fe$_{2-y}$Se$_2$, (TlK)$_x$Fe$_y$Se$_2$, (TlRb)$_x$Fe$_y$Se$_2$ *etc.*, which formed a new group of superconducting iron-based materials without toxic As. In this paper the recent progress in synthesis of 122-like iron-selenide systems and in experimental research of their properties is reviewed. Available theoretical data on electronic, magnetic, and elastic properties of this newest group of superconducting materials are also discussed.

# 1. Introduction

The discovery in 2008 of superconductivity with transition temperature $T_C$ ~ 26 K in fluorine-doped LaFeAsO [1] and subsequent finding of a rich series of related so-called iron-based superconductors (SCs), with maximal $T_C$ ~ 55 K for SmFeAsO$_{1-x}$F$_x$, CeFeAsO$_{1-x}$F$_x$, PrFeAsO$_{1-x}$F$_x$, and NdFeAsO$_{1-x}$F$_x$ [2-4], aroused much interest in a new family of high-temperature superconductors (HTSCs) alternative to the earlier known group of cuprate ceramic HTSC materials [5].

Today, several groups of iron-based SCs are known. The majority of them are so-called iron-pnictide phases (Fe-$Pn$, where $Pn$ are pnictogens). These materials can be categorized into five major classes: 111 (such as $A$FeAs, where $A$ are alkali metals [6]), 122 ($B$Fe$_2Pn_2$, where $B$ are alkali-earth metals [7]), 1111 (such as aforementioned LaFeAsO$_{1-x}$F$_x$, SmFeAsO$_{1-x}$F$_x$), and more exotic multi-component 32225 (such as Sr$_3$Sc$_2$Fe$_2$As$_2$O$_5$ [8]) or 21113 phases (such as Sr$_2$$M$Fe$Pn$O$_3$, where $M$ are Sc, Ti, V *etc.* [9,10], and their homologs such as Ca$_{n+2}$(Al,Ti)$_n$Fe$_2$As$_2$O$_z$; n = 2, 3 or 4 [11]).

The crystal structures of these Fe-$Pn$ materials include 2D-like [Fe$_2Pn_2$] blocks, which are formed by edge-shared tetrahedra {Fe$Pn_4$}. In turn, these [Fe$_2Pn_2$] blocks are separated by $A$ or $B$ atomic sheets (for 111 and 122-like phases, respectively) or by [$Ln$O], [$A_3M_2$O$_5$], [$A_2M$O$_3$] or [Ca$_{n+1}$(Sc,Ti)$_n$O$_y$] blocks for more complex 1111, 32225 or 21113 phases. Note that the electronic bands in the window around the Fermi level are formed mainly by the states of these [Fe$_2$As$_2$] blocks and play the main role in superconductivity, whereas the $A$ and $B$ atomic sheets or oxide blocks serve as "charge reservoirs". An exciting feature of Fe-$Pn$ SCs is that the aforementioned 111, 122, 1111 non-doped parent compounds are located on the border of magnetic instability and commonly exhibit temperature-dependent structural and magnetic phase transitions with the formation of antiferromagnetic (AFM) spin ordering. These compounds have high chemical flexibility to a large variety of constituent elements, together with high structural flexibility. Superconductivity emerges either by hole or electron doping of the parent compounds or under external pressure. Thus, optimum superconductivity for Fe-$Pn$ materials emerges when long-range spin-density-wave (SDW) and concomitant structural transitions are completely suppressed by doping or by pressure, suggesting an unconventional non-BCS pairing mechanism, see reviews [11-20].

Already in 2008 the family of iron-based superconducting materials has been expanded by inclusion of the discovered PbO-type β-FeSe with $T_C$ ~ 8 K [21]. This material became the first in a new class of iron-chalcogenide binary Fe$Ch$ SCs, which are known now as 11 phases, see review [22]. In FeSe, the superconducting transition temperature increases to 15 K with S and Te substitutions [23], but decreases rapidly with electron doping [23–25]. The presence of excess Fe and Se deficiency in FeSe also affects the superconducting properties [26]. With an increase in external pressure to ~ 9 GPa, $T_C$ of FeSe increases to ~ 37 K [22,26–28].

Very recently, the first 122-like ternary iron-chalcogenide superconductor K$_{0.8}$Fe$_2$Se$_2$ with enhanced $T_C$ ~ 31 K was discovered [29]. This phase belongs to the same ThCr$_2$Si$_2$-like structural type as the well known 122-like Fe-$Pn$ SCs such as BaFe$_2$As$_2$, but instead of alkali-earth metals, alkali ions are intercalated between [Fe$_2$Se$_2$] blocks.

Note that except the recently discovered K$_{0.8}$Fe$_2$Se$_2$, a series of ThCr$_2$Si$_2$-type ternary chalcogenides such as $AM_2Ch_2$ (where $A$ = K, Rb, Cs, $M$ = Co, Ni, Cu, and $Ch$ = S, Se) was known earlier [30-35], and some of these phases were considered as magnetic materials [32-34]. However, no data about their superconductivity were available until recently.

The discovery of the first ternary iron-chalcogenide superconductor $K_{0.8}Fe_2Se_2$ [29] has triggered much activity in search for related materials. Very soon (2010-2011), subsequent studies [36-53] confirmed the initial discovery. Moreover, superconductivity ($T_C \sim 28$-$32$ K, which are so far the highest transition temperatures for known Fe chalcogenides) was found in the series of related 122-like Fe-Se phases (termed further as 122FeSe) such as $Cs_xFe_{2-y}Se_2$, $Rb_xFe_{2-y}Se_2$, $(TlK)_xFe_ySe_2$ $(RbK)_xFe_ySe_2$ *etc.*, which have formed today ***the newest group of ternary superconducting iron-based materials without toxic As.***

The available data demonstrate that the 122FeSe systems show a set of distinctive properties as compared with other iron-based SCs.

So, the intercalation of alkali (or $Tl^{1+}$) ions between [$Fe_2Se_2$] blocks is expected to introduce a large number of electrons into the 122FeSe systems. This usually leads to suppression or disappearance of superconductivity as in heavily electron-doped 122-like Fe-As systems, see [11-20]. Therefore, no superconductivity is expected in 122FeSe materials at such a high $T_C$ (over 30 K) considering the high level of electron doping. Next, the available angle-resolved photoemission spectroscopy (ARPES) data [54-57] and band structure calculations [58-63] reveal that such high electron doping in 122FeSe leads to the disappearance of the hole-like Fermi surface sheets. As a result, electron scattering between the hole-like and electron-like bands, which is considered to be important for pairing in Fe-*Pn* SCs, see reviews [11-20], becomes impossible. Besides, the available data reveal that Fe vacancies play a crucial role in electronic and magnetic properties of 122FeSe systems, where superconductivity itself occurs only in Fe-deficient samples [36-53]. This is not typical of Fe-*Pn* SCs. In addition, for Fe-*Pn* materials superconductivity emerges in proximity to the magnetic state, whereas for some 122FeSe systems superconductivity was found to be close to the insulating phase.

The aforementioned (and some other) differences between the well-known Fe-*Pn* SCs and the newly discovered 122FeSe family of SCs have attracted the attention of experimentalists and theorists to new perspectives in the understanding of Fe-based compounds, which will provide hints at the difference and common features between Fe-pnictide and Fe-selenide systems and shed light on the mechanism of superconductivity in Fe-based superconductors.

Here the state-of-the-art results related to the synthesis methods, physical properties (structural, transport, magnetic, electronic, thermodynamic *etc.*), and the progress achieved currently in theoretical understanding of the electronic band structure, magnetic behavior, and elastic properties of the family of the newly discovered ternary 122-type iron-selenide superconductors and some related materials are reviewed as a compendium of the first stage of research on this newest group of superconducting Fe-based systems.

## 2. Synthesis and structure of new ThCr$_2$Si$_2$-type iron-selenide superconducting materials

Already the first discovered superconducting phase $K_xFe_2Se_2$ was synthesized both as polycrystalline and monocrystalline samples [29]. Today, the processes of preparation of high-quality 122FeSe samples are well documented and successfully followed by several groups in China, Switzerland, and Japan and then in Canada and the USA [36-53].

*2.1. Synthesis.*

Usually, for polycrystalline samples, a two-step solid-state reaction method is employed. At the first stage, binary FeSe samples are usually synthesized as precursors utilizing well-known routes, see review [22]. For example, the authors [36] have chosen $FeSe_{0.98}$ as a precursor. This starting material was prepared using high-purity (~ 99.99%, Alfa) powders of Fe and Se, which were mixed and pressed into rods, sealed in evacuated quartz ampoules, and annealed at 700°C during 15h. Further, these species were pressed again into rods in inert atmosphere, sealed in evacuated quartz ampoules, and treated at 700°C over 48h, and were further annealed at 400°C for 36h.

Next, for polycrystalline samples, for example, $K_xFe_2Se_2$ [29], FeSe and K were mixed with appropriate stoichiometry, heated in alumina crucibles, and sealed in quartz tubes with argon. Then they were heated to 973–1023 K, kept for 30 h, and cooled naturally to room temperature. Other possible ways of synthesis of polycrystalline and single-crystalline samples of alkali metal intercalated 122FeSe phases are presented in [37-53].

For the synthesis of single-crystalline $K_xFe_2Se_2$ or $Cs_xFe_2Se_2$ samples [36], a ceramic rod of $FeSe_{0.98}$ was sealed in an evacuated silica ampoule with pure alkali metals: K or Cs, where the quantity of the alkali metal should depend on the desired stoichiometry of the final phases. The ampoules were annealed at 1030°C over 2h for homogenization, the melt was cooled down to 750°C at 6°C/h rate and then cooled down to room temperature with the rate 200°C/h. Well formed black crystalline rods of 7 mm in diameter were obtained, see Fig. 1. These single crystals (with stoichiometry $(K,Cs)_{0.8}Fe_2Se_{1.96}$) could be easily cleaved into plates with flat shiny surfaces [36].

For Tl-containing superconducting materials, at the first stage $TlFe_xSe_2$ (1.5 ≤ x ≤ 2) phases were prepared [40], see also details in earlier works [64-66]. Note that these samples (i) are Fe-deficient, (ii) show AFM ordering with the Neel temperature $T_N$ ~ 365 K, and (iii) a superconducting transition ($T_C^{middle}$ ~ 22.4 K and $T_C^{zero}$ ~ 20 K) was observed in the monocrystalline $TlFe_{1.70}Se_2$ [40].

For more complex compounds, such as $Tl_{0.61}K_{0.39}Fe_{1.76}Se_2$ [40], the aforementioned $TlFe_xSe_2$ species were used as precursors, and then the required phases were prepared as single crystals by adding K in the starting materials. Attempts were undertaken to increase the Fe content further. For this goal, K was added in the starting materials to grow $(Tl,K)Fe_xSe_2$ single crystals[40]. Figure 2 shows the dependence of the composition of $Tl_{0.5}K_zFe_xSe_2$ species on the K content. As the content of K in the starting material decreases, the content of Fe linearly increases in the range $0.3 \leq z \leq 0.45$ and achieves the maximal value 1.83 at z = 0.3.

On the other hand, for the related compound $Tl_{0.58}Rb_{0.42}Fe_{1.72}Se_2$ [43], $Rb_2Se$, $Tl_2Se$, Fe, and Se powders were chosen as precursors, and single crystals were grown by the Bridgeman method.

For the sulfur-doped $K_{0.8}Fe_2Se_2$, the starting precursor $FeSe_{0.8}S_{0.2}$ was prepared from powders of iron, selenium, and sulfur (Fe/Se/S = 1.0/0.8/0.2) at 650°C for 12 hours. Then, using K pieces and $FeSe_{0.8}S_{0.2}$ powder, single crystals $K_{0.8}Fe_2Se_{1.4}S_{0.4}$ were grown from the melt by means of the self-flux method [45]. The first non-superconducting Se-free 122-like phase $K_{0.8}Fe_2S_2$ was prepared [67] also by the self-flux method from FeS and K. The same self-flux method was used [51] for synthesis of a series of solid solutions $K_xFe_{2-y}Se_{2-z}S_z$, where z = 0; 0.4; 0.8; 1.2; 1.6; 2.0.

Finally, note that the synthesis of the discussed Fe-Se(S) 122 phases (Table 1) is more simple than that of Fe-As compounds and requires no strict handling and care owing to the absence of toxic arsenic.

*2.2. Crystal structure.*

The crystal structure of 122-like Fe-Se phases is depicted in Fig. 3 together with typical powder XRD and single crystal XRD data for $K_xFe_2Se_2$ and $Cs_xFe_2Se_2$ [39]. These 122-like materials adopt a tetragonal $ThCr_2Si_2$-type structure (space group *I*4/*mmm*; #139). Here, the Fe atoms form a square lattice, whereas the Se atoms are located at the apical sites of the tetrahedrons $\{FeSe_4\}$; in turn, these tetrahedrons form quasi-two-dimensional (2D) blocks $[Fe_2Se_2]$. In general, the structure of $AFe_2Se_2$ can be schematically described as a stacking of *A* sheets and $[Fe_2Se_2]$ blocks in the sequence: …$[Fe_2Se_2]/A/[Fe_2Se_2]/A/[Fe_2Se_2]$… as shown in Fig. 3. This atomic stacking is clearly visible on the high-resolution transmission-electron-microscopy (TEM) image [68] of $KFe_{2-x}Se_2$, Fig. 4. The atomic positions are *A*: 2*a* (0, 0, 0), Fe: 4*d* (0, ½, ¼), and Se: 4*e* (0, 0, $z_{Se}$), where $z_{Se}$ is the so-called internal coordinate. The available lattice parameters (*a*, *c*, and *c/a*) for the synthesized 122FeSe samples and some related materials are presented in Table 1.

Note that in the sequence $KFe_2Se_2 \rightarrow RbFe_2Se_2 \rightarrow CsFe_2Se_2$, the lattice parameters grow as the radii of K, Rb, Cs ions increase: $R(K) = 1.51$ Å $< R(Rb) = 1.63$ Å $< R(Cs) = 1.78$ Å. Besides, the actual compositions of the crystals sufficiently affect the lattice parameters, and therefore the effect of anisotropic deformation of the lattice should be expected in the presence of Fe or Se vacancies. For $K_{0.8}Fe_2Se_{1.4}S_{0.4}$, both *a* and *c* parameters become smaller than those for $K_{0.8}Fe_2Se_2$ indicating an efficient substitution of S ($R^{atom} = 1.22$ Å) for Se ($R^{atom} = 1.60$ Å) [45]. For Se-free $K_{0.8}Fe_2S_2$ [67], the parameters *a* and *c* become even smaller, see Table 1. In turn, for the solid solutions $K_xFe_{2-y}Se_{2-z}S_z$, (z = 0 ÷ 2), the lattice parameters decrease gradually with an increase in the S content and follow approximately the Vegard's law [51].

The compositions of all the synthesized 122 Fe-Se phases are far from the ideal stoichiometry $AFe_2Se_2$. Thus, one of the interesting issues is the vacancy ordering effect. TEM technique was applied [68] to study structural inhomogeneity and defect structures in the superconducting $KFe_{2-x}Se_2$. The results [68] reveal a variety of microstructure phenomena in this material. In particular, the presence of a superstructure was established in $KFe_{1.5}Se_2$ with the modulation wave vector $q_1$ = (1/5, 3/5, 0), which can be interpreted as the Fe-vacancy ordering within the *ab* plane. TEM studies show that the presence of atomic vacancies and their ordering can be important factors affecting the transport properties of these materials.

Additional evidences about vacancy ordering effects for two monocrystals, $K_xFe_{2-y}Se_2$, have been obtained within XRD experiments [44], where the tetragonal superstructure was found, see Fig. 5. The ordering effects of cationic ($K^+$ and $Fe^{2+}$) vacancies were obtained within XRD for $K_{0.93}Fe_{1.52}Se_2$ and $K_{0.86}Fe_{1.56}Se_2$ species in Ref. [69], where these materials were declared as tetrahedral vacancy-ordered derivatives of the $ThCr_2Si_2$ structure. The presented [69] tetragonal cell is a five-fold expansion in the *ab* plane of the unit cell of the simple square Fe lattice found in other iron-based SCs, with a low occupancy of one Fe site surrounded by a square of four fully occupied Fe sites. Thus, the proposed structural motif is a 4/1 occupied/vacancy ratio with an ideal composition of blocks $[Fe_{1.6}Se_2]$. In this case, the "ideal" composition $K_{0.8+x}Fe_{1.6-x/2}Se_2$ corresponds to the $Fe^{2+}$ valence state, while deviations from this relationship are indicative of electron or hole doping. It was proposed also that in the presence of two different Fe sites, selective metal doping with formation of mixed-cation materials can arise owing to various coordination numbers of the two Fe sites. On the other hand, the potassium vacancies were found positionally disordered.

The neutron powder diffraction spectra for $K_{0.8}Fe_2Se_{1.6}$ were measured [47] at various temperatures from 11 to 580 K. In the proposed structure (space group *I*4/*m*, # 87) the Fe vacancies occupy the *4d* sites, whereas the Fe sites *16i* are fully occupied, and the order-disorder transition of Fe vacancy occurs at T ~ 578 K.

A similar study for $Cs_{0.83}Fe_{1.71}Se_2$ within neutron and x-ray powder and single crystal synchrotron diffraction shows [48] a superstructure with an ordered pattern of Fe vacancies corresponding to the space groups $P4_2/n$ and *I*4/*m*. The defined cell (5 times bigger than the unit cell) corresponds to the propagation vector star of transformation ***A*** = 2***a*** + ***b***, ***B*** = - ***a*** + 2***b***, ***C*** = ***c***. Recently, further evidences for Fe-vacancy ordering structures were obtained in neutron diffraction studies for a set of 122FeSe phases with the common composition $A_2Fe_4Se_5$, where *A* are Cs, K, (Tl,Rb), and (Tl,K) [53].

Further systematic investigations of superconductivity in the 122FeSe systems as depending on the vacancy content and vacancy ordering seem very desirable.

The systematic studies of the Fe-As materials allow one to suggest some correlations between the structural parameters and the transition temperature $T_C$. So, a $T_C$ dependence of the orthorhombic distortion $\delta = (c-a)/(c+a)$ was reported for some 122 Fe-As phases, see review [20]. For the discussed 122 Fe-Se phases, such correlations remain still unclear. Besides, the correlations between $T_C$ and the bond angle of As-Fe-As was found, where the formation of the regular {FeAs$_4$} tetrahedron is important for enhanced $T_C$ [70]. For $K_xFe_2Se_2$, the Se-Fe-Se angle was found ~ 110.9° [29], *i.e.* it is closer to the ideal tetrahedron angle (109.47°) as compared to that of the binary FeSe [22]. For $K_{0.92}Fe_{1.54}Se_2$, the Se-Fe-Se bond angles vary from 104.23º to 114.29º [44]. It was pointed out [44] that the inclusion of alkali ions and the presence of lattice vacancies within the $ThCr_2Si_2$-type structure allow a considerable relaxation of the {$FeSe_4$} tetrahedrons as compared with the same for the binary Fe(Te,Se,S) systems, and the resulting decrease in the distortions of the Se-Fe-Se angles could be an important factor for the understanding of the increase in $T_C$ in ternary selenides, see also [67].

Some preliminary data have been obtained also for the relationship of the so-called anion height ($\Delta z_a$) versus $T_C$. For Fe-pnictogen systems it was established that the pnictogen height from the Fe plane ($\Delta z_{Pn}$) is a good guideline for higher $T_C$ [71]. The theoretical background is that the parameter $\Delta z_{Pn}$ may be a switch between high-$T_C$ nodeless and low-$T_C$ nodal pairings for the Fe-based SCs [71]. For these systems, the dependence ($T_C$ versus $\Delta z_{Pn}$) has a clear maximum at about $\Delta z_{Pn} = 1.38$ Å [72].

Figure 6 depicts the anion height $\Delta z_a$ versus $T_C$ for some of the synthesized $K_xFe_2Se_2$ and $Cs_xFe_2Se_2$ samples [36] in comparison with binary FeSe(Te) and Fe-As systems. Note that these 122 Fe-Se phases follow the universal trend. However, some differences in the properties of these materials (see below) require further explanations of this correlation. The possibility of tuning conductivity between a superconducting metal and a magnetic insulatorby adjusting the lattice parameters and the anion height was reported in [67] using $K_{0.8}Fe_2S_2$ as an example.

Another interesting circumstance was noted in [46]. For different batches of superconducting and non-superconducting $K_xFe_2Se_2$ and $Rb_xFe_2Se_2$ single crystals, two sets of (00$l$) reflections were observed within XRD experiments, which correspond to two $c$-axis lattice parameters $c_1$ and $c_2$, Table 2. This fact is indicative of inhomogeneous distribution of the intercalated alkali atoms. Besides, superconductivity exist within a limited range of the $c_{1,2}$ parameters, see Table 2. Thus, the occurrence of superconductivity is closely related to the $c$ lattice parameter, which, in turn, depends on the content of intercalated alkali atoms.

## 3. Physical properties

In this Section the key physical properties of the examined 122FeSe phases are presented; among them are the transport properties and then the relations between superconductivity and magnetism and the behavior of 122FeSe phases under pressure are discussed. The available data on electronic, magnetic, and superconducting properties of these materials as obtained within ARPES, NMR, optical and Raman scattering experiments are also discussed.

*3.1. Transport properties.*
As is shown in Fig. 7, the typical normal state resistivity ($R$) of $K_{0.8}Fe_2Se_2$ demonstrates a broad hump around 140-150K and semiconducting characteristics in the higher temperature range [41]. As the temperature decreases, the resistance displays a metallic behavior and drops abruptly at about 31K, which is indicative of superconductivity.

The resistivity (R) depends strongly on the stoichiometry of the examined samples, namely, on the Fe deficiency. Indeed, when the concentration of Fe increases, the above hump diminishes, gradually shifting to higher temperatures and finally vanishes in the sample with the nominal composition $K_{0.8}Fe_{2.3}Se_2$. Similar resistivity humps (in the interval of $T_H \sim$ 110-250 K) were observed for related 122FeSe phases, see [29, 36-39,42,49]. This resistivity anomaly may be caused by a possible semiconductor-to-metal transition [29,41] or by structure or magnetic phase transitions (as was proposed from muon-spin rotation/relaxation (μSR) measurements [73]), which are typical of the 122-like Fe-As SCs, reviews [12-20].

A systematic study for a set of single crystals $K_{0.8}Fe_{2-x}Se_2$, $Rb_{0.8}Fe_{2-x}Se_2$, $Cs_{0.8}Fe_{2-x}Se_2$, $Tl_{0.4}K_{0.3}Fe_{2-x}Se_2$, and $Tl_{0.4}Rb_{0.4}Fe_{2-x}Se_2$ [52] shows that all of the samples display common features. Namely, resistivity shows a broad hump with $T_H \sim$ 70-300 K, Fig. 8. Above $T_H$, resistivity demonstrates a semiconductor-like behavior, whereas a sharp increase in R is observed in range from 512 K to 551 K, indicating a phase transition [52].

The measurements of resistivity in zero field from 1.9 K to 300 K for current along *ab* plane and *c* axis of $K_xFe_ySe_2$ single crystal [74] demonstrate that in the normal state below 300 K, the ratio $R_c$:$R_{ab}$ is about 4-12. For $K_{0.86}Fe_{1.84}Se_{2.02}$, this anisotropy reaches the maximum of 6 around 180 K and decreases to 4 around 300 K [49]. Here, the anisotropy is much smaller than was reported for more complex systems $(Tl,K)_xFe_{2–y}Se_2$ [40] and $(Tl,Rb)_xFe_{2–y}Se_2$ [43], where $R_c$:$R_{ab}$ is about 70-80 and 30-45, respectively.

The Hall coefficient ($R_H$) measured for $K_{0.8}Fe_2Se_2$ is negative over the whole temperature range from 70 K to 250 K, indicating that the conduction carriers are dominated by electrons, and a simple estimation [29] of the carriers concentration gives the value $n \sim 1.76 \times 10^{21}$. The negative value of $R_H$ fixed also for $(Tl,K)Fe_{2-x}Se_2$ [40] shows that the carrier is dominated by electrons, which agrees with the electronic band calculation result for the stoichiometric $TlFe_2Se_2$ [75]. In the examined $(Tl,K)Fe_{2-x}Se_2$, the carrier concentration increases with the content of Fe, the $|R_H|$ value decreasing, Fig. 9. Besides, $|R_H|$ increases sharply at $T$ = 250 K and 123 K owing to electron localization; this may be considered as a typical behavior of an AFM Mott insulator. For a more complex compound, $Tl_{0.58}Rb_{0.42}Fe_{1.72}Se_2$, the value of $R_H$ changes from positive to negative at 214 K [43], which is typical of multi-band systems and means that there are both hole and electron pockets near the Fermi level. Also, for $Rb_xFe_{2-x}Se_2$, the Hall coefficient, positive at a high temperature, gradually decreases with decreasing temperature and becomes negative at low temperatures indicating that hole pockets can exist in the superconducting samples [46]. On the contrary, for the $(Tl,K)Fe_xSe_2$ system [40], only negative $R_H$ values occur from $T_C$ to 350 K showing that the carrier is dominated by electrons.

The thermoelectrical power (TEP) measurements for $K_xFe_ySe_2$ [49,50] as a function of temperature demonstrate a negative TEP sign showing that electron-like carriers are dominant, Fig. 10. The origin of the observed local minimum and maximum of the TEP curve in the interval from 100 to 200 K remains unclear, but can be possibly associated [49] with the multiband structure of the system. Within thermal transport measurements for the $K_{0.65}Fe_{1.41}Se_2$ single crystal [50], a set of valuable parameters was estimated: Fermi momentum $k_F \sim 2.6$ nm$^{-1}$, effective mass $m^* \sim 3.4\ m_e$, Fermi velocity $v_F \sim 89$ km/s, and superconducting coherence length $\sim$ 1.8 nm.

Based on the resistivity and magnetic susceptibility ($\chi$) data for $(Tl,K)Fe_xSe_2$ ($1.50 \leq x \leq 1.85$), a phase diagram of this system was proposed (Fig. 11), where three regions with distinct physical properties are visible [40]. So, in the region of $1.3 \leq x < 1.7$, this phase is an AFM insulator, probably, owing to the existence of a super-lattice of Fe-vacancies. The Néel temperature, $T_N$, decreases with increasing Fe content. In the region $1.70 \leq x < 1.78$, a superconducting transition is observed, but the superconducting fraction is small. In the region $1.78 \leq x \leq 1.88$, bulk superconductivity emerges [40].

Recently [51], a phase diagram for solid solutions $K_xFe_{2-y}Se_{2-z}S_z$, ($z = 0 - 2.0$) was proposed, see Fig. 12. It can be seen that sulfur doping leads to lowering of the transition temperature, and superconductivity eventually vanishes at 80% S. Except for the "pure" $K_xFe_{2-y}S_2$, all solid solutions show a resistivity maximum and exhibit a metallic behavior at low temperatures. The temperature of the resistivity hump is found to be not monotonic with the doping level of sulfur, implying that the crossover may be influenced by Fe deficiency [51].

The transition temperatures of 122FeSe materials $T_C$ are summarized in Table 1. The transition width (TW, in the absence of magnetic field) depends on the quality and composition of the samples. So, for the high-quality monocrystalline $K_{0.8}Fe_2Se_2$, the TW value is about 1.3 K [29], for single crystals $K_{0.8}Fe_{2-x}Se_2$, $Rb_{0.8}Fe_{2-x}Se_2$, and $Tl_{0.4}Rb_{0.4}Fe_{2-x}Se_2$ the TW are smaller than 1 K [52], whereas for the phase with the nominal composition $Tl_{0.5}K_{0.34}Fe_2Se_2$ $T_C^{midpoint} = 25.1$ K and $T_C^{onset} < 2.1$ K [40].

A typical temperature dependence of resistivity in various magnetic fields $H$ both parallel and perpendicular to the $c$-axis is plotted in Fig. 13 for $K_{0.86}Fe_2Se_{1.82}$ and $Cs_{0.86}Fe_{1.66}Se_2$ [39]. $T_C$ is suppressed gradually and the TW is broadened with an increase in the magnetic field. The difference in superconductivity in $H//c$ and $H//ab$ fields is clearly visible.

Within the Werthamer-Helfand-Hohenberg (WHH) theory, the upper critical field at T = 0K ($H_{c2}(0)$) was estimated as 208 T and 67 T for $K_{0.86}Fe_2Se_{1.82}$ and 126 T and 43 T for $Cs_{0.86}Fe_{1.66}Se_2$, when $H$ is parallel and perpendicular to the $ab$ plane, respectively. The anisotropy $H_{c2}^{ab}/H_{c2}^{c}$ is about 3.1 for $K_{0.86}Fe_2Se_{1.82}$ and 2.9 for $Cs_{0.86}Fe_{1.66}Se_2$ [39]. Comparable anisotropy ($\sim 3$) was found for $Rb_{0.8}Fe_2Se_2$ [42], and a larger value of $H_{c2}^{ab}/H_{c2}^{c} \sim 5$ was measured for $Tl_{0.58}Rb_{0.42}Fe_{1.72}Se_2$ [43]. These values of $H_{c2}^{ab}/H_{c2}^{c}$ for 122FeSe materials are comparable with those for Fe-As systems, but smaller than those for cuprate SCs.

*3.2. Superconductivity and magnetism.*

The close proximity of AFM and superconductivity in the well studied 122-like Fe-As systems [12-20] has drawn the attention of scientists to the search for similar relations in the new ternary 122FeSe SCs. For the binary Fe-Se(Te) systems, a direct correlation of superconductivity and spin fluctuations was demonstrated in pressure experiments, when the transition temperature for FeSe increased from 8 K to 37 K under high pressures, accompanied by spin fluctuation enhancement [22].

Originally, the $K_xFe_2Se_2$ phase was determined as a Pauli paramagnet [29], according to the magnetization data, where in the temperature range from room temperature to $T_C^{onset}$, the zero-field cooled (ZFC) and field cooled (FC) magnetization curves were found to be flat and temperature independent. In addition, for $K_xFe_{2-y}Se_2$, no evidences for a magnetically ordered phase were found in $Se^{77}$ NMR experiments [76] because NMR lines are very sharp and exhibit a paramagnetic behavior.

On the other hand, some experimental evidences were obtained recently about the relation of superconductivity and magnetism for these materials. So, weak ferromagnetic or AFM fluctuations were assumed for these systems from other NMR data [77-79]. The μSR measurements for $Cs_{0.8}Fe_2Se_{1.96}$ [73] indicate that the system is magnetic below $T_N \sim 478$ K, and there is a microscopic coexistence of superconducting and magnetic phases. According to [40], the evolution from a superconducting state to an insulating AFM state for these systems depends on the Fe content, as is illustrated in Fig. 11. Besides, electron spin resonance (ESR) spectra for $K_{0.8}Fe_2Se_{1.4}S_{0.4}$ [45] show evidence for strong spin fluctuations at temperatures above $T_C$. In addition, a novel resonance signal below $T_C$ was interpreted as an indicator of a novel magnetic state, which evolves in the superconducting matrix, or as occurring owing to the local moment of Fe ions. According to the first-principles calculations, the ground state of $AFe_2Se_2$ crystals ($A$ = K, Cs, or Tl) is a so-called bi-collinear or stripe-like AFM state, see Sec. 4.

The magnetic susceptibility measurements (at $H$ = 5 T in the temperature range to 600 K, Fig. 14) for single crystals $K_{0.8}Fe_{2-x}Se_2$, $Rb_{0.8}Fe_{2-x}Se_2$, $Cs_{0.8}Fe_{2-x}Se_2$, $Tl_{0.4}K_{0.3}Fe_{2-x}Se_2$, and $Tl_{0.4}Rb_{0.4}Fe_{2-x}Se_2$ [52] indicate AFM antiferromagnetic transitions; the corresponding $T_N$ are listed in Table 3.

Very intriguing data about the coexistence of long-range magnetic order and superconductivity were obtained recently within neutron diffraction experiments [47] for the so-called charge-balanced (where iron atoms are in $Fe^{2+}$ charge states) crystal $K_{0.8}Fe_{1.6}Se_2$ and for a set of 122FeSe phases with common composition $A_2Fe_4Se_5$, where $A$ are Cs, K, (Tl,Rb) and (Tl,K) [53]. The AFM order and Fe vacancy order were determined in the $a\sqrt{5}\times\sqrt{5}\times 1$ cell. The obtained ordered magnetic moment of Fe (3.31 $\mu_B$) at 11K is the largest value among all iron pnictide and chalcogenide SCs, and magnetic transition also occurs at a record high value $T_N \sim 559$ K, see also Table 3. The magnetic unit cell depicted in Fig. 15 contains a pair of [FeSe] blocks related by inversion symmetry and the magnetic ordering vector is $\mathbf{q}_m$ = (101). The Fe magnetic moments form a collinear AFM structure with the $c$-axis being the most easy magnetic axis, and this magnetic order coexists with superconductivity. The authors [52] pointed out that that the realization of superconductivity in such a strong magnetic environment may open up a new avenue to magnetic HTSC. Further evidences for Fe-vacancy ordering structures and magnetic ordering were obtained in the aforementioned neutron diffraction studies for $A_2Fe_4Se_5$ phases, where $A$ are Cs, K, (Tl,Rb), and (Tl,K) [53]. Here, both $AFe_2Se_2$ of the $ThCr_2Si_2$ structure and $AFe_{1.5}Se_2$ ($A_2Fe_4Se_5$) of the $\sqrt{2}\times\sqrt{2}\times1$ supercell structure have been proposed as the structure framework for the new 122FeSe SCs; some of their properties are summarized in Table 3.

Thus, the available data are quite inconsistent, and further studies of relations between magnetic and superconducting effects are necessary, which may shed light on the mechanism of superconductivity in these systems. Note also that the $K_{0.8}Fe_{2-x}S_2$ phase was found insulating and showing a magnetic order possibly of a glassy nature below 32K [67]

### 3.3. Specific heat.

Low-temperature specific heat (SH) was measured recently for the superconducting $K_xFe_{2-y}Se_2$ with $T_C \sim 32K$ [80]. From the SH data in the low-temperature region (below 8K) fitted for $H = 0$ T and 9 T as $C(T,H) = \gamma(H)T + \beta T^3 + \eta T^5$ (where $\gamma(H)T$ is the residual SH coefficient in the magnetic field $H$ and $\beta T^3 + \eta T^5$ is the phonon part of heat capacity) it was obtained that $\gamma(0) \sim 0.394$ mJ/mol·K$^2$, $\beta \sim 1.018$ mJ/mol·K$^4$, and $\eta \sim 0.003$ mJ/mol·K$^6$. From these data, the Debye temperature was estimated to be $\Theta_D \sim 212$ K. Thus, $\Theta_D$ is relatively small as compared to 122 Fe-As-based superconductors [20].

In Fig. 16, the SH data plotted as $C/T$ *versus* $T$ demonstrate a rather sharp SH anomaly at $T_C$, which is weakened and shifted to lower temperatures, when a magnetic field is applied. The height of this SH anomaly $\Delta C/T$ near $T_C$ is estimated to be about 11.6 mJ/mol·K$^2$, which is smaller than for related Fe-As materials [20]. This rather sharp SH anomaly allows us to suggest [80] that the superconducting transition in 122FeSe materials can be described quite well by the critical mean field theory.

Next, from the low-temperature part of the SH, the field induced enhancement of the SH was obtained, which exhibits a near-linear field dependence indicating a nodeless gap [80]. From the SH data in the finite-temperature region, the value of the normal-state electron SH coefficient $\gamma_n = 5.8$ mJ/mol·K$^2$ was evaluated. The value $\Delta C/\gamma_n T|_{Tc} = 1.93$ shows that the system possesses a strong coupling regime, since the weak coupling limit (within BCS theory) is 1.43. Note also that the SH data were analyzed [80] within the *s*-wave scaling law and were found to roughly obey this law, which is indicative of an *s*-wave gap.

On the other hand, for $K_{0.86}Fe_{1.84}Se_{2.02}$, the obtained value $\Delta C/\gamma_n T = 1.33$ [49] is close to the weak coupling BCS regime.

### 3.4. Pressure effects.

It is well known that pressure is one of the most effective ways (along with doping effects) of inducting superconductivity of iron-based SCs, reviews [12-20]. For these materials external pressure tends to destroy the magnetic transition in the undoped parent phases, whereas $T_C$ increases with increasing pressure for underdoped species, remains approximately constant for the optimal doping level, and decreases linearly in the overdoped limit [81,82].

For binary iron chalcogenides FeSe(Te), low-temperature superconductivity ($T_C \sim 8$ K) is found under ambient conditions, whereas $T_C$ increases rapidly to ~37 K under pressure [22].

It is quite natural to examine the pressure effects for the new 122FeSe SCs, for which $T_C$ is almost four times higher than that of the binary FeSe(Te) at ambient pressure.

Meanwhile, such experiments (at relatively low pressures to $P < 12$ GPa) have been performed for $K_xFe_2Se_2$ [83-85] and $Cs_xFe_2Se_2$ [85,86] single crystals. For $K_{0.8}Fe_{1.7}Se_2$, the pressure derivation of the transition temperature, $dT_C/dP$, is negative, and at $P \sim 9.2$ GPa superconductivity is completely suppressed, Fig. 17. No structural phase transition was observed over the pressure range examined in [83]. In addition, the correlation between the pressure dependences of $T_C$ and the resistance hump temperature ($T_H$) was detected; probably, this fact may be useful [83] for the understanding of the pairing mechanism in these superconducting materials. For $Cs_{0.8}Fe_2Se_2$, superconductivity is completely suppressed at $P \sim 8$ GPa [86].

The data for $K_{0.8}Fe_2Se_2$ [84] reveal that the $T_C^{zero}$ was suppressed with increasing pressure, but $T_C^{onset} = 33$ K at ambient pressure was enhanced by applying external pressure and reached 36.6K at $P \sim 2$ GPa. This difference can be due to the quality of samples or inhomogeneity of applied pressure [85].

Comparison of the behavior of two batches of single crystals $K_xFe_2Se_2$ with different transition temperatures ($K_{0.85}Fe_2Se_{1.80}$ with $T_C^{onset} = 32.7$ K *versus* $K_{0.86}Fe_2Se_{1.82}$ with $T_C^{onset} = 31.1$ K, which will be denoted further as KFSI and KFSII, respectively) under pressure reveals [85] that for KFSI $T_C$ decreases, whereas for KFSII as well as for $Cs_{0.86}Fe_{1.66}Se_2$ a dome-like dependence of $T_C(P)$ is observed, Fig. 18. So, for KFSII $T_C^{onset}$ increases with pressure and reaches the maximum value 32.7 K at $P \sim 0.48$ GPa, and then it monotonically decreases. Similarly, for $Cs_{0.86}Fe_{1.66}Se_2$ $dT_C/dP \sim 1.3$K/GPa in the interval of $P < 0.82$ GPa, where $T_C^{onset}$ increases to 31.1 K at P = 0.82 GPa from 30 K at ambient pressure and gradually decreases as the pressure grows further. Such a different behavior can be due to a different content of the alkali metal, as well as to the presence of Fe or Se lattice vacancies, which form some "doping level" of each system. Then, if the optimal doping level has been achieved (as for KFSI with $T_C \sim 32.7$ K), the external pressure tends to decrease $T_C$, whereas in other cases a non-monotonic $T_C(P)$ dependence will be observed [85].

Besides, the examination of the pressure cycle for $Cs_{0.8}Fe_2Se_2$ [86] (*i.e.* upon increasing and decreasing the external pressure) demonstrates that $T_C$ is reversible, but the results [86] allow one to suggest that superconductivity is not related to the resistance hump.

Thus, the available experiments [83-86-i14] indicate that unlike binary Fe*Ch* SCs, the transition temperatures of 122FeSe materials cannot be further optimized by applying high pressures.

*3.5. ARPES and NMR experiments.*

Angle-resolved photoemission spectroscopy (ARPES) was successfully used [54-57,87] for experimental investigations of the peculiarities of the low-energy band structure and Fermi surface (FS) for the newest 122FeSe SCs.

For the Fe-deficient single crystal $K_{0.8}Fe_{1.7}Se_2$ ARPES measurements [55] show that one of the distinct features of this material is that the electron-like band crosses the Fermi level around the M point (Fig. 19); a simple parabolic fit allows one to estimate the Fermi velocity as 0.52 eV·Å and the electron mass as $3.5m_o$. On the contrary, no bands cross the Fermi level at the Γ point (Fig. 19); thus, there is no FS pocket observed at the BZ center. These results mean that the FS topology in $K_{0.8}Fe_{1.7}Se_2$ (which contains exclusively electron-like sheets) differs drastically from those for Fe-As SCs [12-20], for which AFM scattering between Γ-centered hole-like and M-centered electron-like FS pockets is a key ingredient for pairing. It is proposed [55] that the FS topology of this Fe-Se based superconductor favors (π,π) inter-FS scattering between the electron-like FSs at the M point, Fig. 19.

The electron-type nature of FS pockets was confirmed in independent ARPES experiments both for $K_{0.8}Fe_2Se_2$ and $Cs_{0.8}Fe_2Se_2$ [54]. The high-resolution data taken above and below $T_C$ for $K_{0.8}Fe_2Se_2$ (Fig. 20) indicate that there is no gap at $E_F$ in the normal state, while a gap (ΔE ~ 4 meV) was identified at 10 K and a superconducting gap was found to be ΔE ~ 10.3 meV, *i.e.* ~ $4k_BT_C$. This ratio between ΔE and $T_C$ is similar to other iron-based superconductors [12-20]. By examination of symmetrized lines at various Fermi crossings, an isotropic *s*-wave type gap ΔE ~ 10.3 meV around the M point was found, while a smaller gap arises around the Γ point. As a result, it was concluded [54] that the inter-pocket hopping or FS nesting may not be necessary for unconventional superconductivity in iron-based SCs and the conventional (isotropic) *s*-wave pairing type may be suitable for describing the superconducting state for these systems.

For a more complicated composition $Tl_{0.58}Rb_{0.42}Fe_{1.72}Se_2$ ($T_C$ ~ 32 K) the ARPES measurements revealed quite a distinct FS topology and nearly isotropic superconducting gaps at about 15 meV (near Γ) and 12 meV (near M) [56]. Here, the Fermi surface consists of an electron-like pocket near M (π,π) and two electron-like sheets around the Γ (0,0) point, Fig. 21.

The ARPES data for the related system $Tl_{0.63}K_{0.37}Fe_{1.78}Se_2$ ($T_C^{middle}$ = 29.1K and $T_C^{zero}$ = 27.5K) [57] reveal nearly isotropic superconducting gaps ΔE on two electron-like FS sheets at the M point, while the hole-like band is about ~ 50 meV below the $E_F$, see Fig. 22. The gap ΔE ~ 8.5 meV results in a pairing strength (ΔE/$k_BT_C$ ~ 7) twice stronger than the weak-coupling BCS value. In addition, an electron-like band around Γ was observed. It was noted that strong pairing on these FSs may be affected by the interaction happening beyond the vicinity of $E_F$ (Fig. 22), even at an energy scale as large as the onsite Coulomb interactions [57]. The additional evidences for nodeless superconducting gap in $K_{0.68}Fe_{1.79}Se_2$ (ΔE ~ 9 meV) and $Tl_{0.45}K_{0.34}Fe_{1.84}Se_2$ (ΔE ~ 8 meV) were obtained within the subsequent ARPES experiments [87], where two electron-like FS sheets were established around the zone center Γ(0,0). These and aforementioned [56,57] results are useful for the formation of a universal picture of the FS topology for $A_xFe_{2-y}Se_2$ (A=K, Cs, Rb, Tl).

The electronic states and pairing type for 122 Fe-Se SCs were also discussed using $^{77}$Se NMR data [76-79]. For $K_{0.8}Fe_{2-x}Se_2$ single crystals (with $T_C$ ranging between 29.5 K and 32 K) it was found that the Knight shift ($^{77}K_n$) and the spin-lattice relaxation rate ($1/^{77}T_1$) are isotropic with different field orientations, but increase dramatically with temperature (above 80 K). On the other hand, $^{77}K_n$ decreases below $T_C$; this is an evidence of the singlet pairing symmetry [76,77] and agrees with the isotropic gap observed with ARPES [54]. As to the spin-lattice relaxation rate, the Hebel-Slichter [88] coherence peak is absent below $T_C$, but a sudden drop of $1/^{77}T_1$ occurs. These data allow us to suggest that $K_{0.8}Fe_{2-x}Se_2$ should be probably a multiple *s*-wave pairing or a *d*-wave pairing compound. From the $^{77}$Se NMR data for $K_{0.8}Fe_2Se_2$ [76,79] it was also believed that both $s_\pm$-wave and *d*-wave pairing types with strong-coupling regime seem appropriate candidates.

The studied $K_{0.8}Fe_{2-x}Se_2$ demonstrates [77] a canonical Fermi liquid behavior (testified by the temperature-independent Korringa ratio *S*) without any evidences of SDW fluctuations typical of Fe-As SCs [12-20]; this does not allow us to make any correlations between spin fluctuations and superconductivity. Nevertheless, taking into account the obtained value $S \sim 1.8$, weak ferromagnetic fluctuations may be assumed [77] for this system. In the next work [78], additional evidences (based on the $^{77}$Se NMR experiments) of strongly coupled superconductivity in $K_{0.8}Fe_{2-x}Se_2$ were discussed. In particular, from the measured relaxation rate $1/^{77}T_1$ in the temperature range from $T_C$ to $T_C/2$, the value of the isotropic gap $\sim 3.8\ k_BT_C$ was estimated.

Besides, the $^{77}$Se NMR spectra are split into two peaks with equal intensity at all temperatures in the magnetic field along the *ab* plane. This suggests that K vacancies should have a superstructure, when the local symmetry of the Se sites is lower than the tetragonal four-fold symmetry in the ideal 122-like crystal [76].

Interesting data on the superconducting penetration depth $\lambda_{ab}$ and the carrier density for $K_xFe_{2-y}Se_2$ were also obtained recently from NMR experiments [76]. From the observed increase in the NMR line width, the in-plane London penetration depth $\lambda_{ab}$ was estimated (at $T = 8K$) to be $\sim 190$ nm, which is almost twice as small as that for $LaFeAsO_{0.91}F_{0.09}$ [89]. Next, using the value of $\lambda_{ab}$, the concentration of superconducting electrons $n^* = m^*c^2/4\pi e^2\lambda_{ab}^2$ was estimated (by approximation of the effective mass $m^*$ of superconducting electrons to that of a bare electron): $n^* \sim 1.5 \cdot 10^{-22}$ cm$^{-3}$.

Further evidences of the key role of Fe vacancies in the properties of $K_{0.8}Fe_{2-x}Se_2$ were obtained in optical experiments [90]. The single crystals $K_{0.8}Fe_{1.5}Se_2$ and $K_{0.8}Fe_{1.7}Se_2$ were examined. The results revealed that $K_{0.8}Fe_{1.5}Se_2$ is not a Mott insulator, but behaves as a semiconductor with a small gap at about 30 meV. Unexpectedly abundant infrared-active phonon modes were found, which are attributed to the Fe vacancies. It was concluded that for these materials superconductivity arises in close proximity to the AFM semiconducting state, and only a small amount of carriers is required for transformation of the parent semiconductor to the superconducting state.

For $K_{0.8}Fe_{1.7}Se_2$, the optical measurements [91] reveal a superconducting gap ΔE ~ 3 meV, which is much smaller than that estimated within the ARPES technique, see above.

Finally, in the work [92] the Raman scattering (RS) measurements on a high-quality crystal $K_{0.8}Fe_{1.8}Se$ ($T_C$ ~32 K) at different temperatures were performed for the first time. It was found that there are at least twelve modes in the crystal, which cannot be explained in the framework of the standard 122 structure. It was suggested that Fe vacancies form some ordering sub-structures, which affect the local lattice vibrations. Besides, the modes have no obvious Fano-asymmetry; therefore the electron-phonon coupling may be weak in the system.

# 4. Electronic band structure, magnetic and some other properties of new ThCr$_2$Si$_2$-type iron-selenide systems from first-principles calculations

Appreciable progress has been achieved currently in theoretical understanding of the electronic band structure and some properties of the discussed 122FeSe materials.

*4.1. Electronic band structure and Fermi surface.*

The first notion of the band structure of ideal stoichiometric 122FeSe systems follows from non-magnetic DFT calculations for $KFe_2Se_2$ [58-61] and $CsFe_2Se_2$ [59,61]. Such calculations can serve as a starting point for the understanding of the basic features of the electronic structure of these materials.

Figure 23 shows the near-Fermi band structure of $KFe_2Se_2$ along the selected high-symmetry lines within the first Brillouin zone of the tetragonal crystal for two sets of structural parameters: taken from experiment [29] for the composition $K_{0.8}Fe_2Se_2$ ($KFS^{exp}$) and as obtained from the full structural optimization of this phase over the lattice parameters and the atomic positions including the internal coordinate $z_{Se}$ ($KFS^{calc}$). It is seen that there is no gap at $E_F$, which indicates that the electrical conductivity of this phase should be metallic-like. The band structure is strongly anisotropic, showing that the conductivity for the single crystal $KFe_2Se_2$ should be highly anisotropic too.

The total density of states (DOS) and atomic and orbital decomposed partial DOSs (PDOSs) for $KFe_2Se_2$ (Fig. 24) resemble those of 122-like Fe-*Pn* systems (see [12-20]) and exhibit typical characteristics of these layered structures. However, for $KFe_2Se_2$ the Se 4*p* states occur between -6.5 eV and -3.5 eV with respect to the Fermi level ($E_F$= 0 eV) and are separated from the near-Fermi bands by a gap. In turn, the bands between -2.4 eV and $E_F$ are mainly of the Fe 3*d* character. Besides, the contributions from the valence states of K to the occupied bands are very small, while the K 4*s* states are located above the $E_F$ in the interval from +2.0 eV to +6.0 eV. Thus, potassium atoms in $KFe_2Se_2$ are in the form of cations close to $K^{1+}$ and provide charge transfer into conducting blocks $[Fe_2Se_2]^{n-}$.

One of the most important features of the electronic band structure are the low-dispersive bands. For the family of 122 Fe-$Pn$ materials these bands cross the Fermi level and are responsible for the Fermi surface (FS) topology [12-20]. The considered 122FeSe phases have an increased number of valence electrons ($nve$ = 29 $e$ per formula unit for "ideal" stoichiometry $M$Fe$_2$Se$_2$) as compared with isostructural FeAs-based materials (for example, BaFe$_2$As$_2$, where $nve$ = 28 $e$ per formula unit). As a result, the Fermi level shifts to the upper Fe $d$-like bands with higher $k_z$ dispersion, whereas the Fe $3d_{zx(y)}$ and Fe $3d_{x^2-y^2}$ bands (which are responsible for the formation of hole pockets in the 122 Fe$Pn$ phases) for KFe$_2$Se$_2$ are fully occupied.

The Fermi surface of KFe$_2$Se$_2$ (Fig. 25) consists of two quasi-two-dimensional (2D) electron-like sheets in the corners of the Brillouin zone and of closed disconnected electron-like pockets (around $Z$), which are extended along the $k_z$ direction acquiring a pronounced conical shape [58,60,61] - instead of cylinder-like hole-like sheets for 122 Fe-$Pn$ materials [12-20]. As a result, the Fermi surface nesting effect in KFe$_2$Se$_2$ is absent. Besides, two electronic-like sheets around $M$ points in KFe$_2$Se$_2$ are much more symmetric than those in the 122 Fe-$Pn$ phases and are independent of the doping level [61].

Thus, based on the results [58,60,61] it is possible to conclude that the electronic state of the *stoichiometric* KFe$_2$Se$_2$ is far from those required for the iron-based high-temperature SCs, and the main factor responsible for the experimentally observed superconductivity for this material should be the cation deficiency, *i.e.* the hole doping effect.

Using the simple rigid band model (shifting downwards the Fermi level) it was demonstrated [58,60,61] that hole doping (K deficiency) leads to the transformation of the aforementioned closed $Z$ centered pocket into cylinder-like hole sheets typical of 122 Fe-$Pn$ SCs, Fig. 25. Besides, the required cylinder-like topology of the hole-like sheets for KFS$^{calc}$ is achieved only at a high level of K deficiency (x = 0.6), whereas for KFS$^{exp}$ (with experimental lattice parameters for K$_{0.8}$Fe$_2$Se$_2$) the cylinder-like sheets are formed already at x = 0.8 [58]. Thus, it was concluded [58] that the tuning of the electronic system of the $M$Fe$_2$Se$_2$ materials may be achieved by a joint effect owing to structural relaxation and hole doping. It is possible to assert also that the structural factor is responsible for the modification of the band topology, whereas the doping level determines their filling.

Based on DFT calculations of KFe$_2$Se$_2$ and CsFe$_2$Se$_2$, simple BCS-like estimations of $T_C$ for these materials were performed [63]. Using the calculated values of the total DOS at the Fermi level N(E$_F$) = 3.94 states/eV·cell for KFe$_2$Se$_2$ and N(E$_F$) = 3.6 states/eV·cell for CsFe$_2$Se$_2$ and the Debye frequency $\omega_D$ = 350K, the authors [e4] derived the coupling constant $\lambda$ = 0.21 eV [94]. Then employing the BCS expression for $T_C = 1.14\omega_D e^{-2/\lambda N(E_F)}$, the values 34 K (KFe$_2$Se$_2$) and 28.6 K (KFe$_2$Se$_2$) were obtained. Besides, it was noted that (i) hole doping leads to an increase in N(E$_F$), and the same estimations at the 60% doping level give $T_C$ = 57 K for K$_x$Fe$_2$Se$_2$ and $T_C$ = 52 K for Cs$_x$Fe$_2$Se$_2$, showing the potential role of doping; (ii) the values of $T_C$ correlate with the values of N(E$_F$), and (iii) the above estimations do not necessarily imply electron-phonon pairing, as $\omega_D$ may just denote the average frequency of any other possible Boson responsible for pairing interaction (e.g. spin fluctuations) [63].

*4.2. Chemical bonding.*

The peculiarities of inter-atomic bonding in 122 Fe-Se phases were discussed in Ref. [93] using the *stoichiometric* $KFe_2Se_2$ as an example. From the DOSs picture (Fig. 24) it was concluded that the states between -2.7 eV and the Fermi level are mainly of the Fe $3d$ character and form metallic-like Fe-Fe bonds. In turn, the Se $4p$ - Fe $3d$ states, which are localized between -6.5 eV and -3.7 eV with respect to the Fermi level, are strongly hybridized and are responsible for the covalent bonding Fe-Se inside [$Fe_2Se_2$] blocks. These covalent bonds are clearly visible in Fig. 26, where the charge density map for $KFe_2Se_2$ is depicted. Weak direct Se-Se inter-block interactions are also seen.

Within the simplified ionic model, the charge transfer ($1e$) from $K^{1+}$ sheets to [$Fe_2Se_2$]$^{1-}$ blocks should be assumed if we use the usual oxidation numbers of atoms: $K^{1+}$, $Fe^{2+}$, and $Se^{2-}$. Within Bader [95] analysis, the effective charges were calculated [93]: $\Delta Q(K) = +0.78e$, $\Delta Q(Fe) = +0.50e$, and $\Delta Q(Se) = -0.89e$. The results show that $KFe_2Se_2$ is a partly ionic compound, and charges are transferred from K and Fe to Se; besides, charge transfer occurs also from $K^{0.78+}$ sheets to [$Fe_2Se_2$]$^{0.78-}$ blocks. In summary, the data [93] reveal that the inter-atomic bonding in this system is highly anisotropic and includes ionic, covalent, and metallic contributions, where mixed covalent-ionic Fe-Se and metallic-like Fe-Fe bonds take place inside [$Fe_2Se_2$] blocks, whereas the inter-blocks (..K/[$Fe_2Se_2$]/K/[$Fe_2Se_2$]..) bonding is basically of the ionic type (owing to K → [$Fe_2Se_2$] charge transfer) with additional weak covalent Se-Se bonding.

Additional research is required for a more precise definition of the charge transfer in cation-deficient species such as charge-balanced $K_{0.8}Fe_{1.6}Se_2$.

*4.3. Magnetic structure. Role of atomic vacancies.*

A universal characteristic of 122 Fe-*Pn* phases is a collinear antiferromagnetic (AFM) order (so-called antiferromagnetic stripe spin ordering) below the tetragonal orthorhombic structural transition [12-20]. For the binary Fe-Se system, the AFM stripe state [96] or the bi-collinear AFM order [97-99] were shown to be the lowest energy states.

Two main types of magnetic ordering for 122FeSe phases are predicted theoretically by examining $KFe_2Se_2$ [60], $AFe_2Se_2$ ($A$ = K, Cs, or Tl) [59], and Fe-deficient $TlFe_{2-x}Se_2$ [62,63]. In case of the stoichiometric $KFe_2Se_2$ [60], it was found from comparison of NM, ferromagnetic (FM), AFM, and stripe-like (collinear) AFM states that the ground state of this system is a stripe-like AFM configuration with ~ 2.26 $\mu_B$ magnetic moments (MMs) on Fe atoms. Within more detailed calculations for $AFe_2Se_2$, four possible magnetic structures with FM, square Neel AFM, collinear AFM, and bi-collinear AFM orders were considered [59]. The MMs for each Fe atom were found to be about 2.4 – 3.0 $\mu_B$, varying weakly for various magnetically ordered states, but the ground state for both $KFe_2Se_2$ and $CsFe_2Se_2$ has a bi-collinear AFM order (Fig. 27).

To quantify the magnetic interactions, a frustrated $J_1$-$J_2$-$J_3$ Heisenberg model was employed, and the values of the nearest, next-nearest, and next next nearest neighbor couplings $J_1$, $J_2$, and $J_3$ were calculated [59]. It is known that the bi-collinear AFM state is lower in energy than the collinear AFM state if $J_3 > J_2/2$ and $J_2 > J_1/2$. The obtained $J_{1-3}$ values are: for $CsFe_2Se_2$ $J_1 = -11.78$ meV/$S^2$, $J_2 = 19.75$ meV/$S^2$, and $J_3 = 11.75$ meV/$S^2$; for $KFe_2Se_2$ $J_1 = -22.3$ meV/$S^2$, $J_2 = 15.6$ meV/$S^2$, and $J_3 = 14.6$ meV/$S^2$. Both $CsFe_2Se_2$ and $KFe_2Se_2$ are found [59] to be quasi-2D bi-collinear antiferromagnetic semimetals with MMs of ~ 2.8 $\mu_B$ around each Fe atom.

On the other hand, recent calculations [100] of charge-balanced $K_{0.8}Fe_{1.6}Se_2$ reveal that this phase is a quasi-2D blocked checkerboard AFM semiconductor with a gap of 0.6 eV and with large (~3.4 $\mu_B$) magnetic moments of Fe atoms. These conclusions agree well with neutron diffraction experiments [47]. The underlying mechanism of formation of such ground state of $K_{0.8}Fe_{1.6}Se_2$ is the chemical-bonding-driven lattice distortion, when in the presence of the Fe vacancies such a lattice distortion occurs, that every four Fe atoms around a Se atom are contracted together into a block to make stronger Fe-Fe and Fe-Se bonding, which leads to energy gain [100].

According to the aforementioned experiments, see Sec. 3, the (Tl,K)-intercalated compounds $(Tl,K)Fe_xSe_2$ (1.3 < x < 1.7) are antiferromagnetic insulators, especially $TlFe_{1.5}Se_2$ with one-quarter ordered Fe-vacancy, whereas earlier calculations [75] showed that the "ideal" $TlFe_2Se_2$ behaves as a metal.

This issue was considered theoretically in Refs. [62,63,101]. So, the two-band model was applied to $(K,Tl)Fe_{1.5}Se_2$. This system was found to be a Mott insulator arising from correlation effects enhanced by Fe vacancies [101].

The *ab-initio* calculations for $TlFe_{1.5}Se_2$ [63] for some possible stacking configurations of Fe-deficient sheets show that the energy differences among different inter-blocks magnetic orderings are very small (< 1 meV/Fe) and therefore this kind of magnetic coupling is negligible. For the Fe-vacancy ordered orthorhombic $TlFe_{1.5}Se_2$ superstructure with stripe-like AFM spin ordering, the electronic structure calculations [63] demonstrate that the band gap is vanishingly small within the conventional generalized gradient correction approximation - GGA [102]. On the other hand, using the Hubbard on-site energy *U*-dependent correction (GGA+*U*), the band gap was found to increase with the value of *U*, see Fig. 28. Thus, it is possible to suggest that the experimentally observed gap in $TlFe_{2-x}Se_2$ is not due to the SDW ordering itself, but originates from the correlations effects.

The magnetic ordering effects in the interplay with Fe-vacancies order were examined in more detail for $TlFe_{1.5}Se_2$ and $AFe_{1.5}Se_2$ (*A* = K, Rb, and Cs) [62]. For these phases, the structure with rhombus-ordered Fe vacancies is more stable *versus* the square-like Fe-vacancies ordering, Fig. 29. Further calculations for this Fe-ordering type with four possible spin configurations (Fig. 29) revealed that the ground state of $TlFe_{1.5}Se_2$ is an AFM semiconductor with a band gap of 94 meV. $KFe_{1.5}Se_2$ is also an AFM semiconductor like $TlFe_{1.5}Se_2$, whereas $RbFe_{1.5}Se_2$ and $CsFe_{1.5}Se_2$ behave as zero-gap semiconductors or AFM semi-metals. The magnetic moment of Fe atoms were found to be about 2.5-2.8 $\mu_B$, varying slightly in the above four magnetically ordered states [62].

The *ab initio* calculations of (K,Tl)Fe$_{1.6}$Se$_2$ [103] reveal that the ground state of this crystal is a checkerboard of AFM coupled Fe$_4$ blocks with local magnetic moments of about 2.8 $\mu_B$ per Fe atom. The obtained gapless electronic structure is of a 3D-type with a unique Fermi surface topology, thus the formation of a superstructure is crucial for the physical properties of this phase.

*4.4. Pairing symmetries.*

To date, several microscopic models were proposed to explain the pairing symmetries for the new 122FeSe SCs [104,105]. So, assuming the insulating state of Tl(K)Fe$_{1.5}$Se$_2$ as a Mott insulator and using the two-orbital Hubbard model in the strong coupling limit, the authors [105] argue that the spin-singlet *s*-wave superconductivity is favorable. However, according to the model based on the functional renormalization group (FRG) method [106], the leading and sub-leading pairing symmetries should be of nodeless $d_{x^2-y^2}$ and nodal extended *s* types, respectively. The authors [107,108] find that the pairing state in these systems is most likely to have the *d*-wave symmetry.

Generally speaking, besides these models, several different pairing symmetries based on the experimental data and *ab-initio* calculations are anticipated: so-called *s*-, $s_\pm$-, and *d*-type pairing. In spite of available arguments for each pairing type, there is still no consensus about the pairing mechanism in these SCs.

*4.5. Elastic properties.*

The elastic properties of 122-like Fe-Se phases seem important for possible applications of these new superconducting materials and can provide some hints on the mechanism of superconductivity in these SCs. Indeed, the elastic constants can be linked to such important physical parameters of superconductors as the Debye temperature $\Theta_D$ and the electron-phonon coupling constant $\lambda$ [109]. Some other correlations between the superconducting critical temperature $T_C$ and the mechanical parameters have been also discussed [110].

Using the *ab initio* method, the authors [93] predicted the elastic properties for KFe$_2$Se$_2$ as a parent phase for 122FeSe SCs. The values of six independent elastic constants ($C_{ij}$) for the tetragonal KFe$_2$Se$_2$ ($C_{11}$ = 146 GPa, $C_{12}$ = 57 GPa, $C_{13}$ = 41 GPa, $C_{33}$ = 64 GPa, $C_{44}$ = 32 GPa, and $C_{66}$ = 68 GPa) are positive and satisfy the well-known Born's criteria [111] for mechanically stable tetragonal crystals: $C_{11}$ > 0, $C_{33}$ > 0, $C_{44}$ > 0, $C_{66}$ > 0, ($C_{11} - C_{12}$) > 0, ($C_{11} + C_{33} - 2C_{13}$) > 0, and {$2(C_{11} + C_{12}) + C_{33} + 4C_{13}$} > 0. Besides, since $C_{33}$ < $C_{11}$ and $C_{44}$ < $C_{66}$, this layered crystal should be more compressible along the *c*-axis and the shear along the (100) plane should be easier relative to the shear along the (001) plane.

Next, the macroscopic elastic parameters for KFe$_2$Se$_2$, namely the bulk (*B*) and the shear (*G*) moduli (in two main approximations: Voigt (V) [112] and Reuss (R) [113]) were evaluated: $B_V$ = 70.4 GPa, $G_V$ = 40.9 GPa and $B_R$ = 57.7 GPa, $G_R$ = 34.9 GPa. Then, within the Voigt-Reuss-Hill (VRH) approximation [114], the corresponding effective moduli for the polycrystalline KFe$_2$Se$_2$ were estimated:

$B_{VRH}$ = 64 GPa, $G_{VRH}$ = 37.9 GPa, as well as the averaged compressibility $\beta_{VRH}$ = $1/B_{VRH}$ = 0.01563 GPa$^{-1}$, the Young modulus $Y_{VRH}$ = $9B_{VRH}G_{VRH}/(3B_{VRH} + G_{VRH})$ = 94.9 GPa, and the Poisson's ratio $v = (3B_{VRH} - 2G_{VRH})/\{2(3B_{VRH} + G_{VRH})\}$ = 0.253. Also, the Lame's constants (physically, the first constant $\lambda$ represents the compressibility of a material while the second constant $\mu$ reflects its shear stiffness) were estimated as $\mu = Y_{VRH}/2(1+v)$ = 37.8 and $\lambda = vY_{VRH}/\{(1+v)(1-2v)\}$ = 38.8.

The data [93] reveal that KFe$_2$Se$_2$ is a relatively soft material. Note also that the elastic moduli of KFe$_2$Se$_2$ are comparable on the whole with the same for other layered Fe-As SC's. For example, according to the available experimental and theoretical data, their bulk moduli are: $B \sim$ 62, 60, 57, and 100÷120 GPa for SrFe$_2$As$_2$, CaFe$_2$As$_2$, LiFeAs, and for some of 1111 FeAs phases such as LaFeAsO or NdFeAsO [115-121], respectively.

The aforementioned elastic parameters allow one to make [93] the following conclusions. For KFe$_2$Se$_2$ $B > G$; this implies that the parameter limiting the mechanical stability of this system is the shear modulus. Since the shear modulus $G$ represents the resistance to plastic deformation while the bulk modulus $B$ represents the resistance to fracture, the value $B/G$ was proposed [122] as a empirical malleability measure of polycrystalline materials: if $B/G > 1.75$, a material behaves in a ductile manner, and *vice versa*, if $B/G < 1.75$, a material demonstrates brittleness. In the present case of KFe$_2$Se$_2$, $B/G \sim 1.69$; thus, this phase should be brittle.

The values of the Poisson ratio for covalent materials are small ($v = 0.1$), whereas for ionic materials a typical value is 0.25 [123]. For KFe$_2$Se$_2$ $v = 0.253$, *i.e.* a considerable ionic contribution to the inter-atomic bonding for this phase should be assumed. Besides, for covalent and ionic materials, typical shear moduli are $G \sim 1.1B$ and $\sim 0.6B$, respectively [123]. For KFe$_2$Se$_2$, the calculated value of $G/B$ is 0.59.

The elastic anisotropy of crystals is an important parameter for engineering science, since it correlates with the possibility of emergence of microcracks in materials. A number of methods were used [93] to estimate the elastic anisotropy of KFe$_2$Se$_2$. So, elastic anisotropy was estimated from coefficients $A$ calculated for every symmetry plane as $A^{(010),(100)} = C_{44}(C_{11} + 2C_{13} + C_{33})/(C_{11}C_{33} + C_{13}^2)$ and $A^{(001)} = 2C_{66}(C_{11} - C_{12})$. Note that for crystals with isotropic elastic properties $A = 1$, while values smaller or greater than unity measure the degree of elastic anisotropy, see for example [124]. In the other method, elastic anisotropy in compressibility ($A_B$) and shear ($A_G$) was estimated (in percents) using the model [125] for polycrystalline materials: $A_B = (B_V - B_R)/(B_V + B_R)$ and $A_G = (G_V - G_R)/(G_V + G_R)$. Finally, the so-called universal anisotropy index [126] defined as $A^U = 5G_V/G_R + B_V/B_R - 6$ was used; for isotropic crystals $A^U = 0$, while deviations of $A^U$ from zero determine the extent of crystal anisotropy. The following values of the aforementioned indexes for KFe$_2$Se$_2$ were obtained: $A^{(010),(100)} = 0.85$; $A^{(001)} = 1.53$; $A_B = 9.9\%$, $A_G = 7.9\%$; and $A^U = 1.08$, indicating considerable elastic anisotropy for this layered material.

In summary, $KFe_2Se_2$ may be characterized as a mechanically stable and relatively soft material, which will adopt considerable elastic anisotropy and demonstrate brittleness [93].

## 5. Summary and outlook

This review is an attempt to give a compendium for the first stage of investigations that have been already performed for the newly discovered family of $ThCr_2Si_2$-type ternary Fe-Se based superconducting materials with a set of novel characteristics within a short span of time - from 2010 to 2011.

Although the family of $ThCr_2Si_2$-type Fe-Se based SCs was discovered only half a year ago, appreciable progress was achieved currently in the development of synthesis routes, understanding of the key physical properties, and in the simulation of electronic, magnetic, and elastic parameters.

So, a representative group of new $ThCr_2Si_2$-type ternary SCs (such as $A_xFe_{2-y}Se_2$, where $A$ are alkali metals, and more complex compositions such as $Tl_xA_zFe_{2-x}Se_2$ or $K_xFe_2Se_{2-y}S_y$, with $T_C \sim 28-33$ K, which are so far the highest transition temperatures for Fe chalcogenides) was prepared and the corresponding synthesis approaches were carefully documented. A lot of new information about structural, transport, electronic, magnetic, elastic, and other properties of these newest materials was obtained both by experimentalists and theorists. A number of similarities and differences as compared with earlier known Fe-$Pn$ SCs were established. For example, the available data reveal that cation vacancies (both in $A$ and Fe sublattices) play a crucial role in the electronic and magnetic properties of 122FeSe systems, where superconductivity itself occurs only in Fe-deficient samples. This is not typical of Fe-$Pn$ SCs. Here, vacancy ordering effects can also play an important role. For some 122FeSe systems, superconductivity was found to be close to the insulating state - unlike the Fe-$Pn$ SCs, where superconductivity emerges in the proximity to the magnetic state. Besides, long-range magnetic order and superconductivity were found to co-exist in the so-called charge-balanced crystal $K_{0.8}Fe_{1.6}Se_2$, for which the magnetic moment of 3.31 $\mu_B$ for Fe is the largest among all iron pnictide and chalcogenide SCs, and magnetic transition occurs at a record high $T_N \sim 559$ K. The realization of superconductivity in such a strong magnetic environment may open up a new avenue to magnetic HTSC [47]. The Fermi surfaces of the 122FeSe systems contain only electron-like sheets - as distinct from the Fe-$Pn$ SCs, where both electron-like and hole-like sheets emerge, which are considered to be important for the pairing mechanism.

The aforementioned (and some other) differences between the well-known Fe-$Pn$ SCs and the newly discovered 122FeSe family of SCs seem very interesting for further understanding of the properties of iron-based SCs, as well as for the development of potential avenues for discovering new high temperature superconductors.

Certainly, despite the aforementioned successes, a lot of issues are left open for future research. So, one of the most interesting topics that deserves further experimental and theoretical attention is the role of chemical tuning of the properties of 122FeSe materials. For example, partial replacement of Fe by other transition metals, such as Co, Ni or Cu will lead to hole or electron doping. This will allow one to control the charge carrier density, which may affect the level of Fe-sites occupancy, and to regulate the cation non-stoichiometry (and the type of vacancy ordering) as one of the key factors of these materials. These substitutions should be also accompanied by a distortion of the tetragonal lattice. Note that very recently the first Co-doped series of $K_{0.8}Fe_{2-x}Co_xSe_2$ samples was synthesized [127]. It was found that the superconductivity in this system is quenched down to ~ 5 K by 0.5 at.% Co doping, and this seems the fastest quenching rate ever-reported.

Of particular interest is the possibility of formation of "mixed" $A_xFe_{2-y}Se_{2-z}As_z$ systems as "intermediate" phases between two groups of iron-based SCs: Fe-*Pn* and Fe-*Ch* ThCr$_2$Si$_2$-type materials. Besides, a lot of other aspects of these materials remain open (for example, detailed magnetic structure, some thermodynamic questions, the interplay of magnetism and superconductivity, pairing symmetry types, vibration properties, surface states etc.).

The author hopes that this review will be useful for further research into layered Fe-Se and related superconducting materials, which seem interesting and challenging systems for providing an additional insight into iron-based HTSCs.

Finally, it is worthy to mention that after submitting this paper, some additional experimental [128-136] and theoretical [137-145] data about discussed materials were appeared on the arXiv.

**Acknowledgments.**
The author acknowledges the support from the RFBR, grants No. 09-03-00946, and No. 10-03-96008.

# TABLES

**Table 1.** Lattice parameters (in Å) and transition temperatures ($T_C$, in K) for synthesized 122FeSe compounds and related materials.

| system | $T_C$ | $a$ | $c$ | $c/a$ | Ref. |
|---|---|---|---|---|---|
| $K_xFe_2Se_2$ | 30 | 3.9136 | 14.0367 | 3.5866 | 29 |
| $K_{0.86}Fe_2Se_{1.82}$ | 31 | 3.8912 | 14.1390 | 3.6336 | 39 |
| $K_{0.8}Fe_2Se_{1.96}$ | 29.5 | 3.9092 | 14.1353 | 3.6159 | 36 |
| $KFe_{2-x}Se_2$ | - | 3.913 | 14.03 | 3.5855 | 68 |
| $K_{0.8}Fe_2Se_2$ | 33 | 3.9034 | 14.165 | 3.6289 | 38 |
| $K_{0.8}Fe_2Se_2$ | ~ 44 [1] | - | 14.15 | - | 41 |
| $K_2Fe_4Se_5$ [6] | 32 | 8.7306 | 14.1128 | 1.6165 | 53 |
| $K_{0.86}Fe_{1.84}Se_{2.02}$ | ~ 30 | 3.8897 | 14.141 | 3.6355 | 49 |
| $K_{0.8}Fe_{1.6}Se_2$ | 32 | 8.6928 [5] | 14.0166 [5] | 1.6124 [5] | 47 |
| $Rb_{0.78}Fe_2Se_{1.78}$ | 32 | 3.925 | 14.5655 | 3.7110 | 37 |
| $Rb_{0.8}Fe_2Se_2$ | ~ 31/28 [2] | - | - | - | 42 |
| $Rb_2Fe_4Se_5$ [6] | 32 | 8.788 | 14.597 | 1.6610 | 53 |
| $Cs_{0.86}Fe_{1.66}Se_2$ | 30 | 3.9618 | 15.285 | 3.8581 | 39 |
| $Cs_{0.8}Fe_2Se_{1.96}$ | 27 | 3.9601 | 15.2846 | 3.8597 | 36 |
| $Cs_{0.83}Fe_{1.71}Se_2$ | - | 3.9614 | 15.2873 | 3.8591 | 48 |
| $Cs_2Fe_4Se_5$ [6] | 29 | 8.865 | 15.289 | 1.7246 | 53 |
| $TlFe_{1.3}Se_2$ | - | 3.90 | 13.89 | 3.5615 | 40 |
| $TlFe_{1.47}Se_2$ | - | 3.90 | 13.95 | 3.5769 | 40 |
| $TlFe_{1.7}Se_2$ | 22.5/20 [3] | - | - | - | 40 |
| $Tl_{0.75}K_{0.25}Fe_{1.85}Se_2$ | 31/28.8 [3] | 3.88 | 14.06 | 3.6237 | 40 |
| $Tl_{0.61}K_{0.39}Fe_{1.76}Se_2$ | 25.1/< 2 [3] | 3.80 | 14.06 | 3.7000 | 40 |
| $(Tl,K)_2Fe_4Se_5$ [6] | 31 | 8.645 | 14.061 | 1.6265 | 53 |
| $Tl_{0.58}Rb_{0.42}Fe_{1.72}Se_2$ | 32/31.4 [2] | 3.896 | 14.303 [4] | 3.6712 | 43 |
| $(Tl,Rb)_2Fe_4Se_5$ [6] | 28 | 8.683 | 14.388 | 1.6570 | 53 |
| $K_{0.8}Fe_2Se_{1.4}S_{0.4}$ | 32.8/31.2 [2] | 3.8560 | 14.0344 | 3.6396 | 45 |
| $K_{0.8}Fe_2Se_{1.6}S_{0.4}$ | 33.2/31.4 [2] | 3.847 | 14.046 | 3.6512 | 51 |
| $K_{0.8}Fe_2Se_{1.2}S_{0.8}$ | 24.6/21.4 [2] | 3.805 | 13.903 | 3.6539 | 51 |
| $K_{0.8}Fe_2Se_{0.8}S_{1.2}$ | 18.2/16.4 [2] | 3.797 | 13.859 | 3.6500 | 51 |
| $K_{0.8}Fe_2Se_{0.4}S_{1.6}$ | non-SC | 3.781 | 13.707 | 3.6252 | 51 |
| $K_{0.8}Fe_2S_2$ | non-SC | 3.753 | 13.569 | 3.6155 | 67 |

[1] "trace of superconductivity" in K- and Fe-deficient sample $K_xFe_{2-y}Se_2$
[2] $T_C^{onset}/T_C^{zero}$
[3] $T_C^{middle}/T_C^{zero}$
[4] An additional series of (00$l$) XRD peaks with $c$ = 14.806 Å was pointed out, *i.e.* there may be a modulation of the structure along the $c$ axis owing to the Fe-vacancy.
[5,6] for Fe-vacancy ordered structures in the tetragonal unit cell (space group $I4/m$, # 87)

**Table 2**. The lattice parameters ($c_1$ and $c_2$, in Å) and transition temperatures ($T_C$, in K) for series of $K_xFe_2Se_2$, and $Rb_xFe_2Se_2$ single crystals [46].

| system | $c_1$ | $c_2$ | $T_C^{zero}$ | $T_C^{onset}$ |
|---|---|---|---|---|
| $K_xFe_2Se_2$(1) * | 14.292 | 14.086 | 31.2 | 31.7 |
| $K_xFe_2Se_2$(2) | 14.282 | 14.062 | 29.2 | 30.8 |
| $K_xFe_2Se_2$(3) | 14.201 | 14.107 | non-SC | |
| $Rb_xFe_2Se_2$(1) | 14.873 | 14.569 | 31.9 | 32.4 |
| $Rb_xFe_2Se_2$(2) | 14.873 | 14.582 | 31.5 | 32.0 |
| $Rb_xFe_2Se_2$(3) | 14.874 | 14.574 | 31.6 | 32.4 |
| $Rb_xFe_2Se_2$(4) | 14.792 | 14.604 | non-SC | |

* (1-4) - examined batches of single crystals

**Table 3.** Neel temperature ($T_N$, in K), temperature of structural transition ($T_S$, in K), magnetic moment at $T \sim 295K$ ($m$, in $\mu_B$), and saturated moment at $T \sim 10K$ ($M$, in $\mu_B$) for some 122FeSe phases as obtained within resistivity and magnetic susceptibility measurements [52] and neutron diffraction studies [53].

| parameter | K * | Rb | Cs | (T,K) | (Tl,Rb) |
|---|---|---|---|---|---|
| $T_N$ | 559 (540) | (534) | 475 (504) | (496) | 513 (500) |
| $T_S$ | (551) | (540) | (525) | (515) | (512) |
| $m$ | 2.76 | 1.9 | 1.9 | 2.1 | 2.0 |
| $M$ | 3.31 | - | 2.3 | 0 | 2.4 |

* The data are given: for 122FeSe phases with common composition $A_2Fe_4Se_5$, where $A$ are Cs, K, (Tl,Rb) and (Tl,K) [53], and in parenthesis - for $K_{0.8}Fe_{2-x}Se_2$, $Rb_{0.8}Fe_{2-x}Se_2$, $Cs_{0.8}Fe_{2-x}Se_2$, $Tl_{0.4}K_{0.3}Fe_{2-x}Se_2$ and $Tl_{0.4}Rb_{0.4}Fe_{2-x}Se_2$ [52]

# F I G U R E S

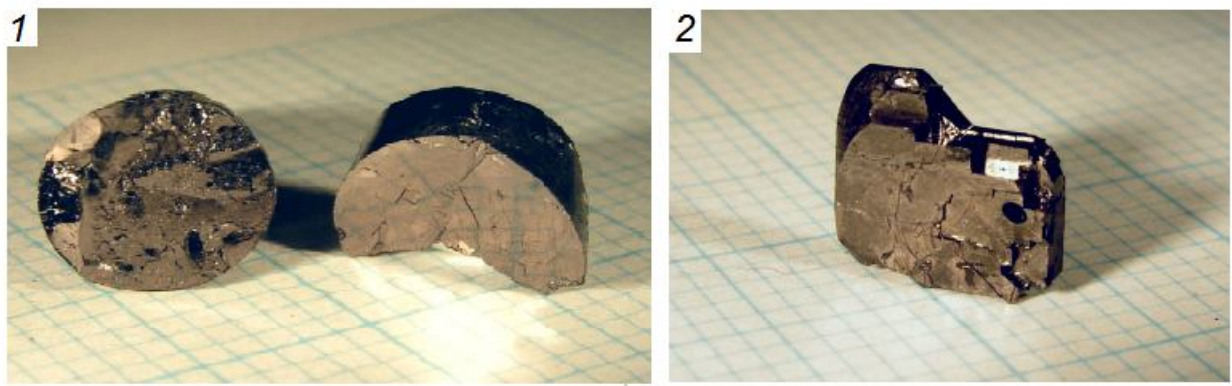

**Fig. 1.** The single crystals of $K_{0.8}Fe_2Se_{1.96}$ (1) and $Cs_{0.8}Fe_2Se_{1.96}$ (2) [36].

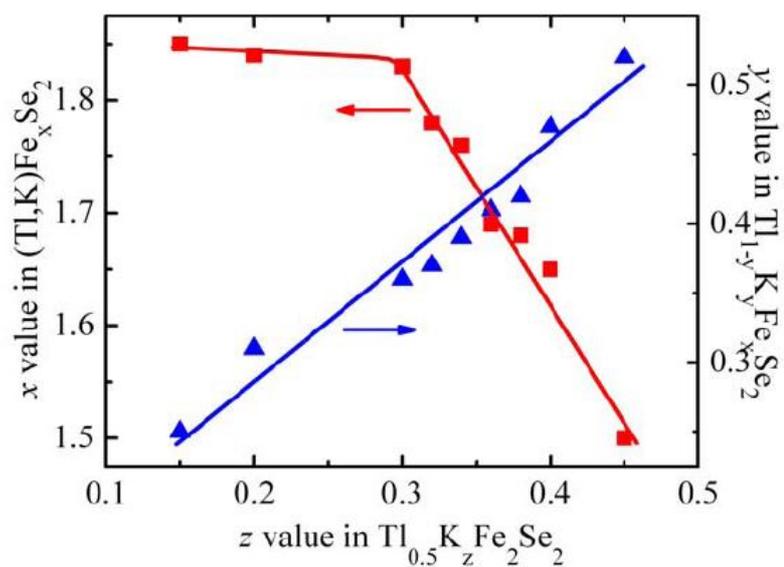

**Fig. 2.** The Fe and K contents in the grown crystal as a function of K content, z value, in the starting materials with the nominal composition $Tl_{0.5}K_zFe_2Se_2$ [40].

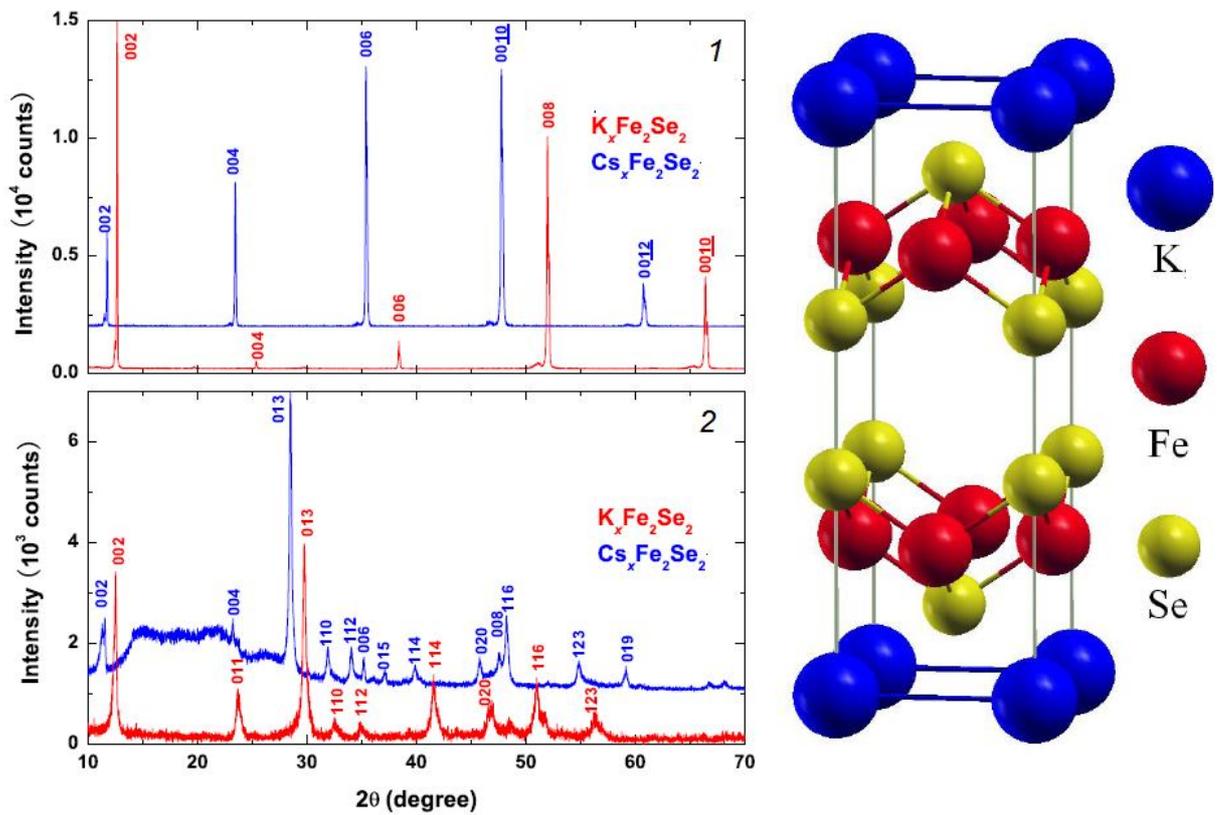

**Fig. 3.** XRD data for (1) single-crystalline and (2) polycrystalline $K_xFe_2Se_2$ and $Cs_xFe_2Se_2$ samples [39]. The crystal structure of $ThCr_2Si_2$-type $KFe_2Se_2$ is also given.

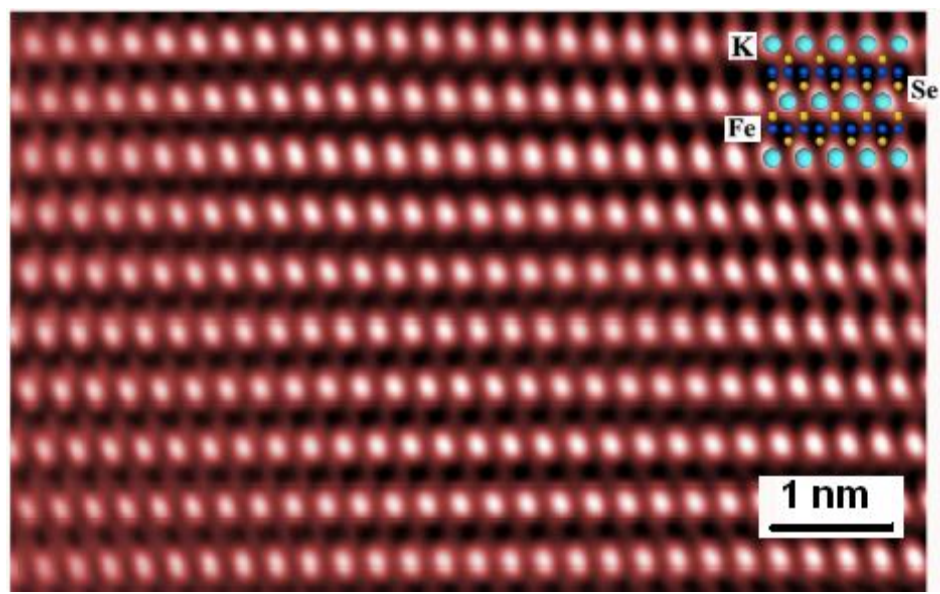

**Fig. 4.** The high-resolution TEM image of $KFe_{2-x}Se_2$ along the [100] zone-axis [68].

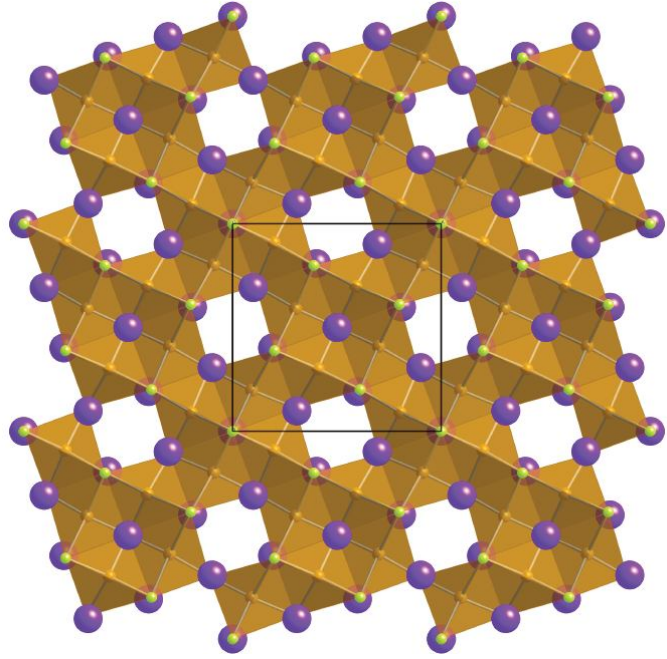

**Fig. 5.** The proposed [44] superstructure of $K_{0.774}Fe_{1.613}Se_2$ from the [001] direction in the √5×√5×1 supercell showing fully occupied Fe sites decorated with ordered vacancy sites.

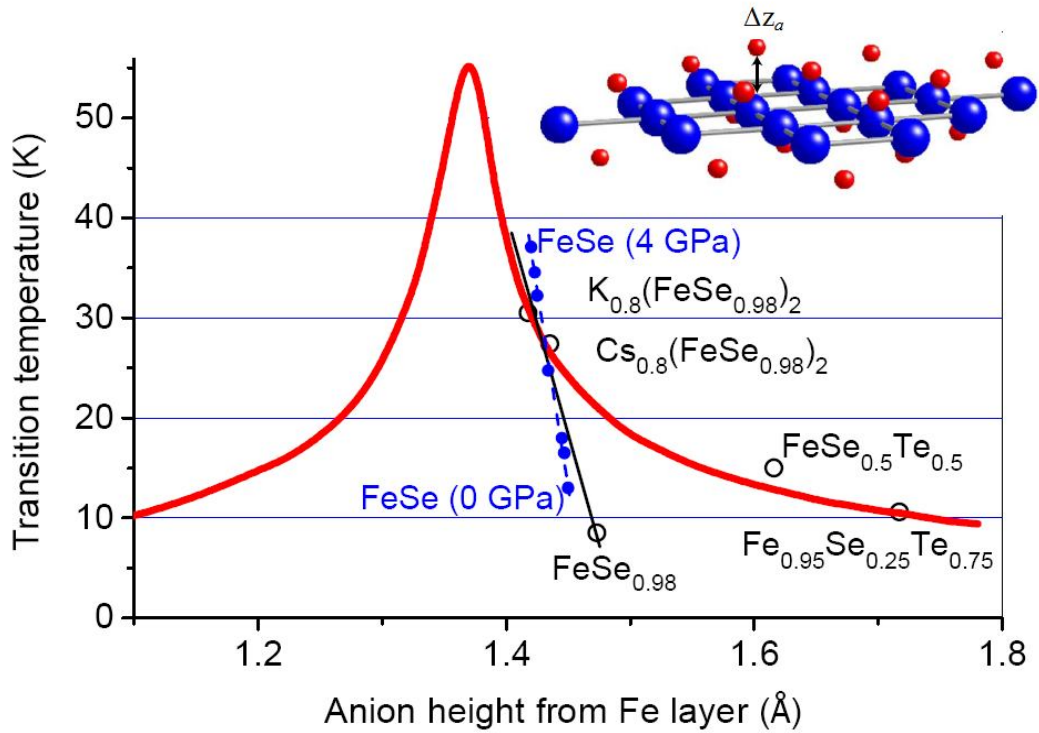

**Fig. 6.** Dependence of $T_C$ *versus* the so-called anion height ($\Delta z_a$, *insert*) for synthesized $(K,Cs)_xFe_2Se_2$ samples as compared with those for the binary FeSe(Te) (full circles) and Fe-As systems (full line) [36].

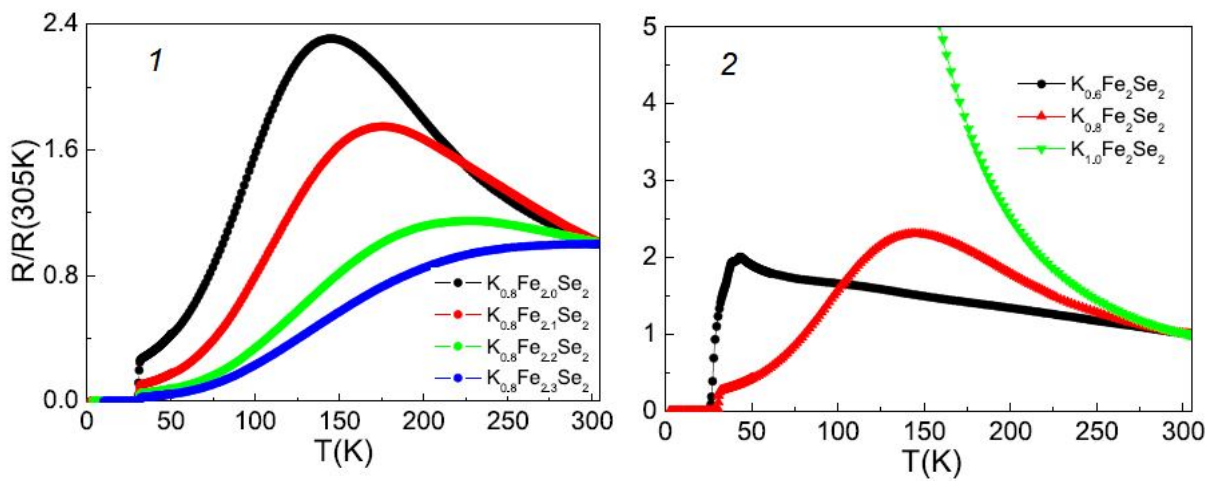

**Fig. 7.** Temperature dependent resistivity of $K_xFe_ySe_2$ samples [41].

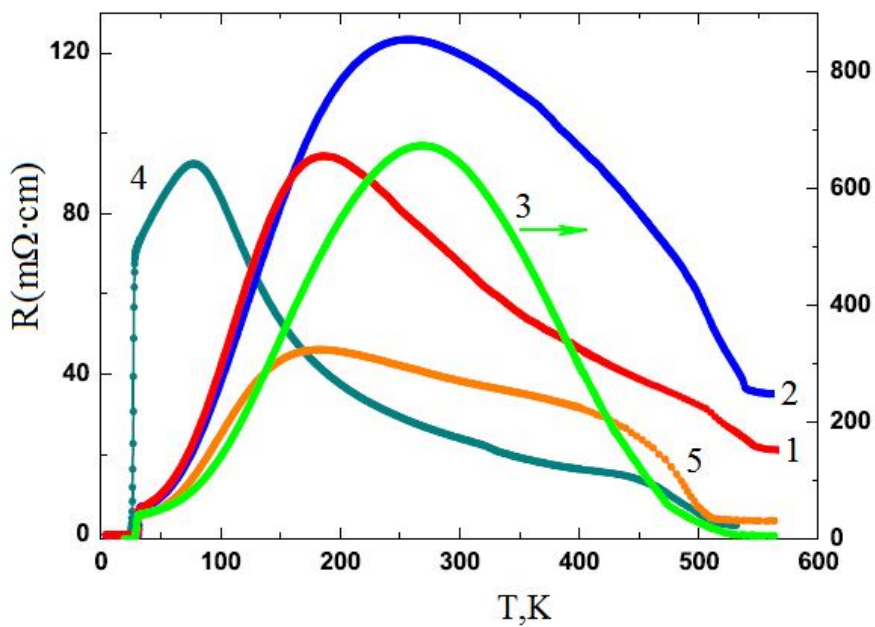

**Fig. 8.** Temperature dependent resistivity of (1). - $K_{0.8}Fe_{2-x}Se_2$, (2). - $Rb_{0.8}Fe_{2-x}Se_2$, (3). - $Cs_{0.8}Fe_{2-x}Se_2$, (4). - $Tl_{0.4}K_{0.3}Fe_{2-x}Se_2$ and (5). - $Tl_{0.4}Rb_{0.4}Fe_{2-x}Se_2$ [52].

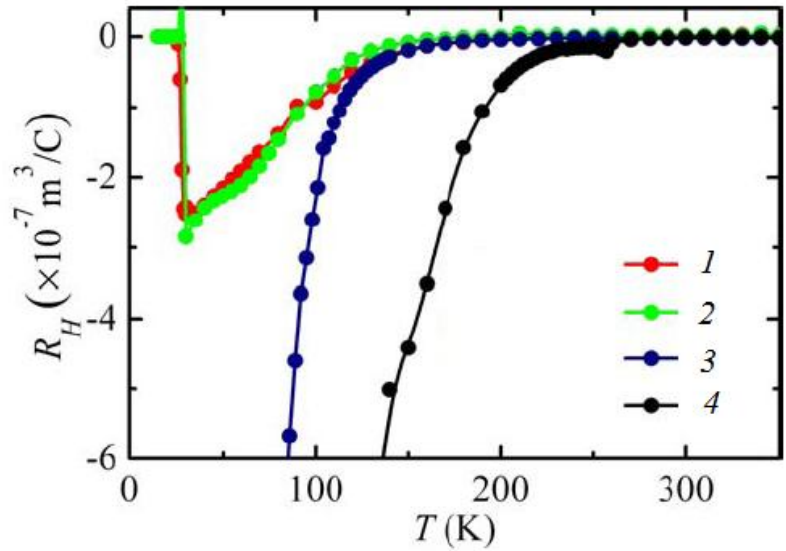

**Fig. 9.** Temperature dependent Hall coefficient $R_H$ of $Tl_{0.75}K_{0.25}Fe_{1.85}Se_2$ (1), $Tl_{0.64}K_{0.36}Fe_{1.83}Se_2$ (2), $Tl_{0.53}K_{0.47}Fe_{1.65}Se_2$ (3), and $Tl_{0.48}K_{0.55}Fe_{1.50}Se_2$ (4) [40].

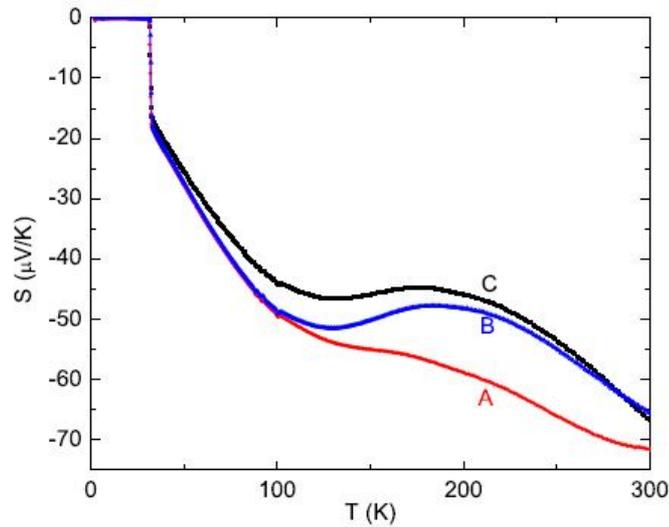

**Fig. 10.** Temperature dependent thermoelectric power of $K_{0.86}Fe_{1.84}Se_2$. Samples A and B use silver paste as contact (contact resistance ~ 1 − 3 Ω). Sample C uses silver wires attached by In-Sn solder as contact (contact resistance ~ 200 Ω) [49].

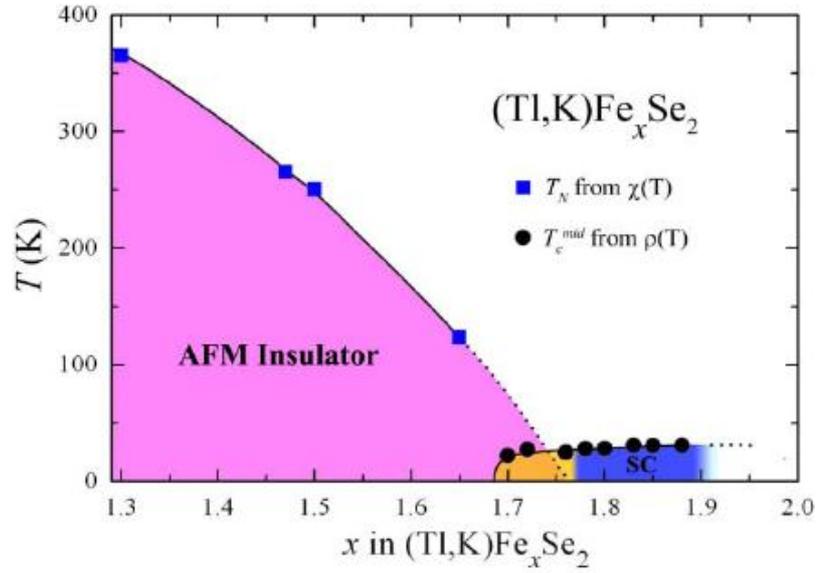

**Fig. 11.** The phase diagram for $(Tl,K)Fe_xSe_2$ ($1.30 \leq x \leq 1.88$) [40].

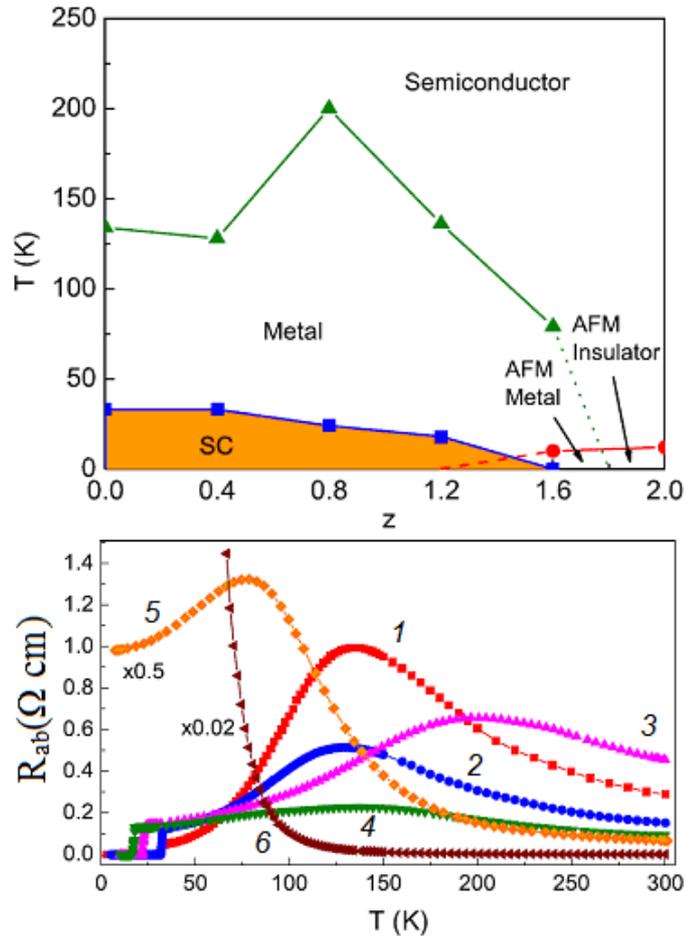

**Fig. 12.** *Top panel*: The phase diagram for $K_xFe_{2-y}Se_{2-z}S_z$ ($z = 0 \div 2$). *Bottom panel*: Temperature dependence of the in-plane resistivity $R_{ab}(T)$ of the $K_xFe_{2-y}Se_{2-z}S_z$ solid solutions at zero field [51].

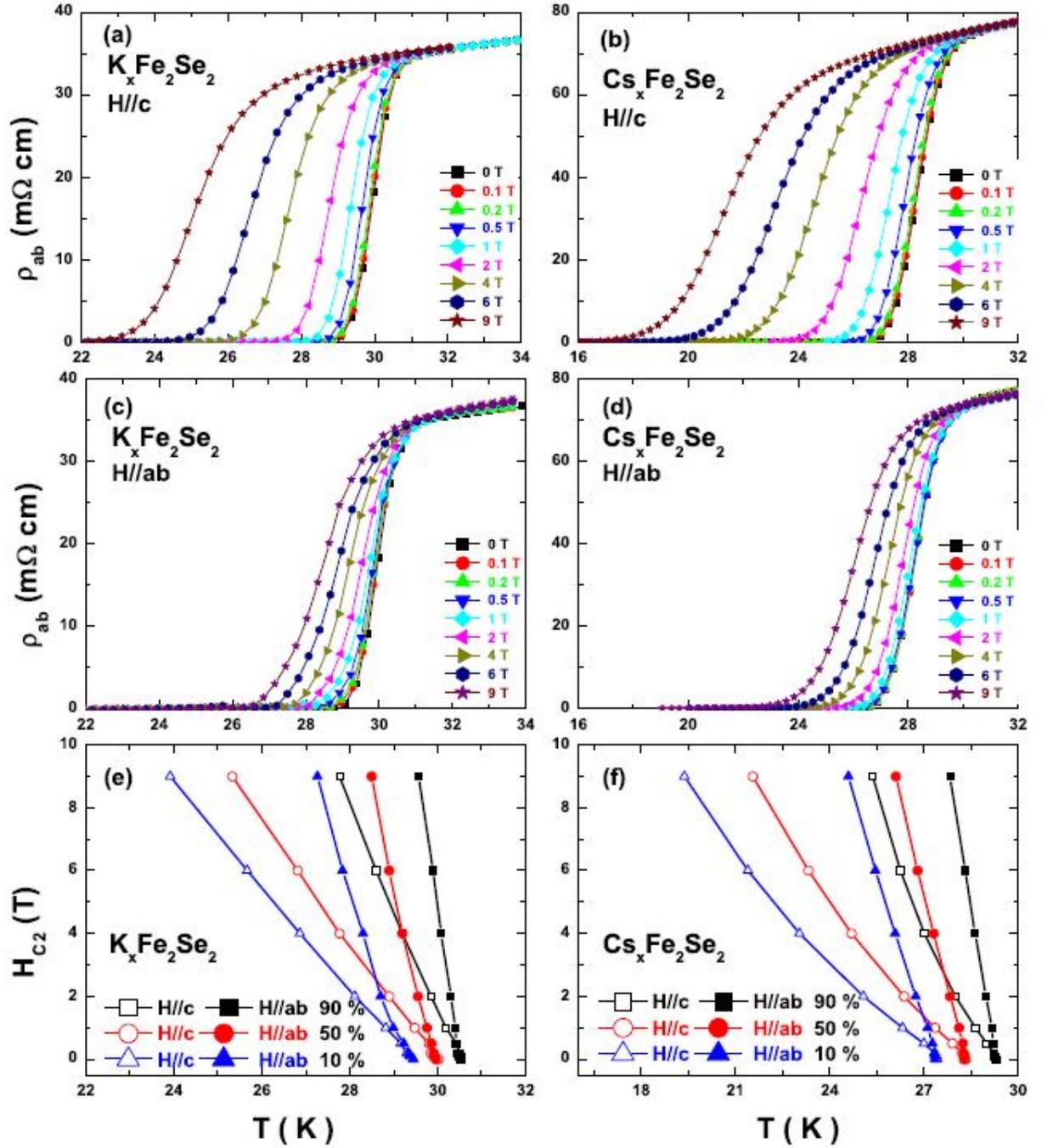

**Fig. 13.** The temperature dependence of resistivity for $K_{0.86}Fe_2Se_{1.82}$ (a,c) and $Cs_{0.86}Fe_{1.66}Se_2$ (b,d) with the magnetic field parallel (a,b) and perpendicular (c,d) to the *c* axis; (e) and (f) show the temperature dependence of $H_{c2}(T)$ for $K_{0.86}Fe_2Se_{1.82}$ and $Cs_{0.86}Fe_{1.66}Se_2$, respectively. Here, $T_C$ is defined as the temperature where resistivity drops by 90%, 50%, and 10% above the superconducting transition [39].

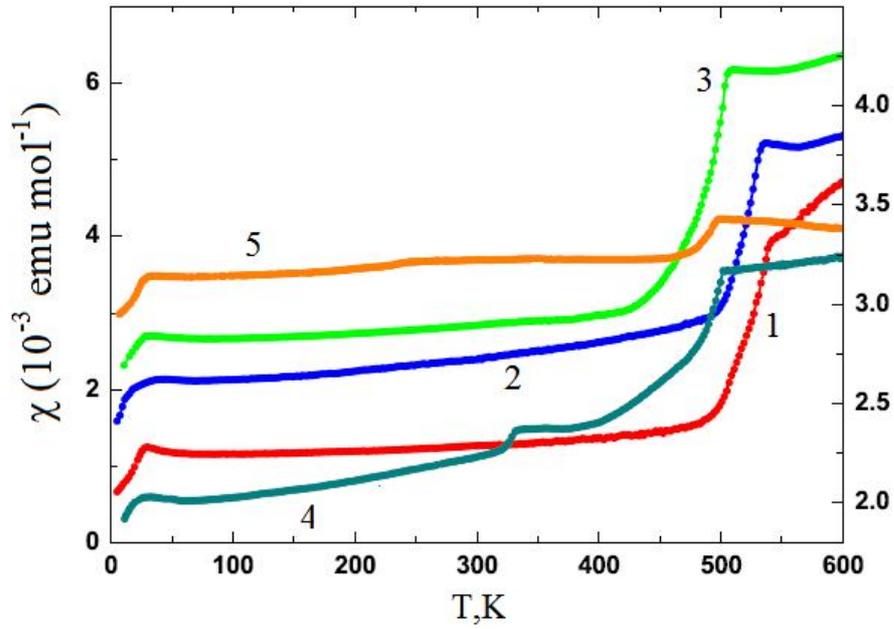

**Fig. 14.** Magnetic susceptibility (at $H$ = 5 T in the temperature range up to 600K) of (1). - $K_{0.8}Fe_{2-x}Se_2$, (2). - $Rb_{0.8}Fe_{2-x}Se_2$, (3). - $Cs_{0.8}Fe_{2-x}Se_2$, (4). - $Tl_{0.4}K_{0.3}Fe_{2-x}Se_2$ and (5). - $Tl_{0.4}Rb_{0.4}Fe_{2-x}Se_2$ [52].

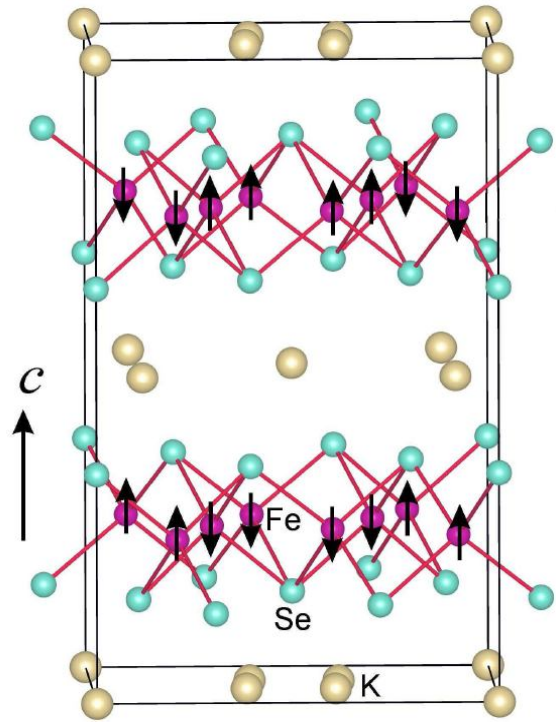

**Fig. 15.** The collinear AFM structure with the $c$-axis as most easy magnetic axis for $K_{0.8}Fe_{1.6}Se_2$ as obtained within neutron diffraction experiments [47,48].

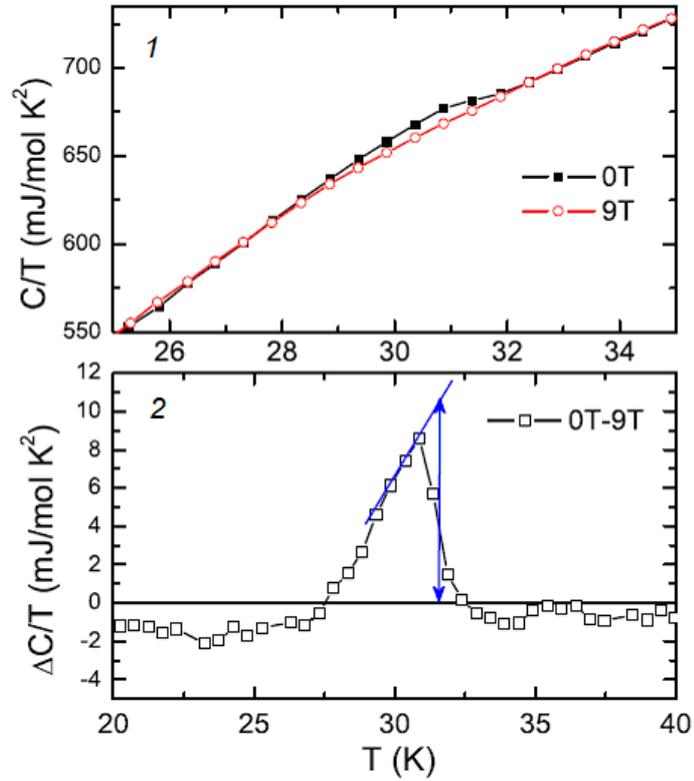

**Fig. 16.** (1). Low-temperature specific heat plotted as *C/T versus* T in magnetic fields $H = 0$ T and 9 T; (2). The difference in the $C/T$ data between 0 T and 9 T. The solid lines show how the height of the $\Delta C/T$ anomaly was determined [80].

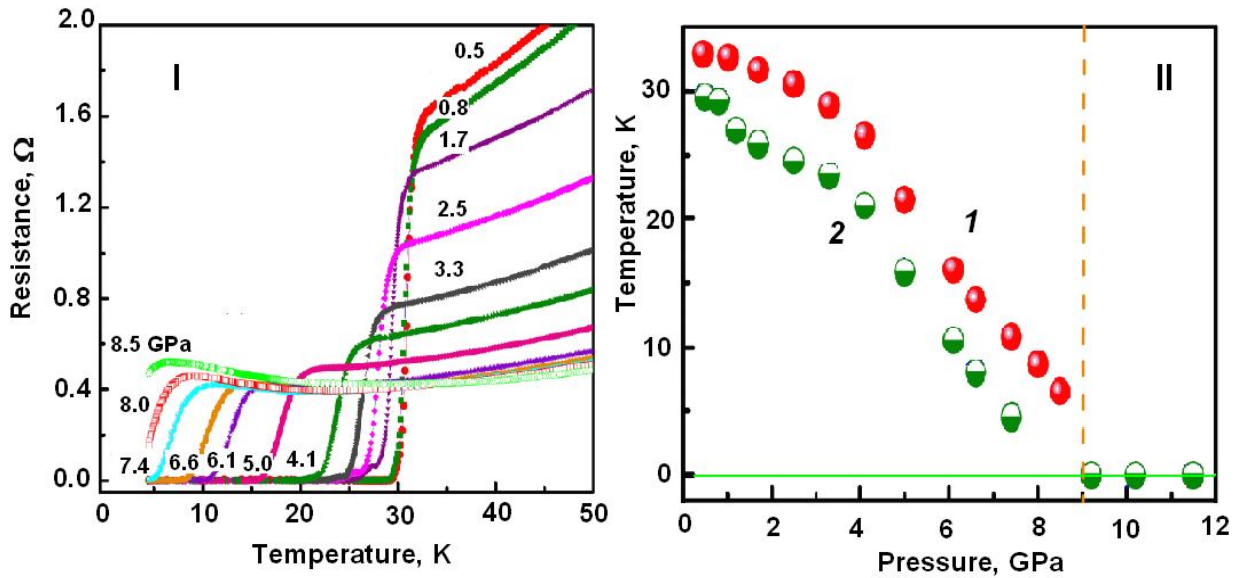

**Fig. 17.** (I) Temperature dependence of electrical resistance at different pressures, and (II) pressure dependence of transition temperature $T_C^{onset}$ (1) and $T_C^{zero}$ (2) for $K_{0.8}Fe_{1.7}Se_2$ single crystal [83].

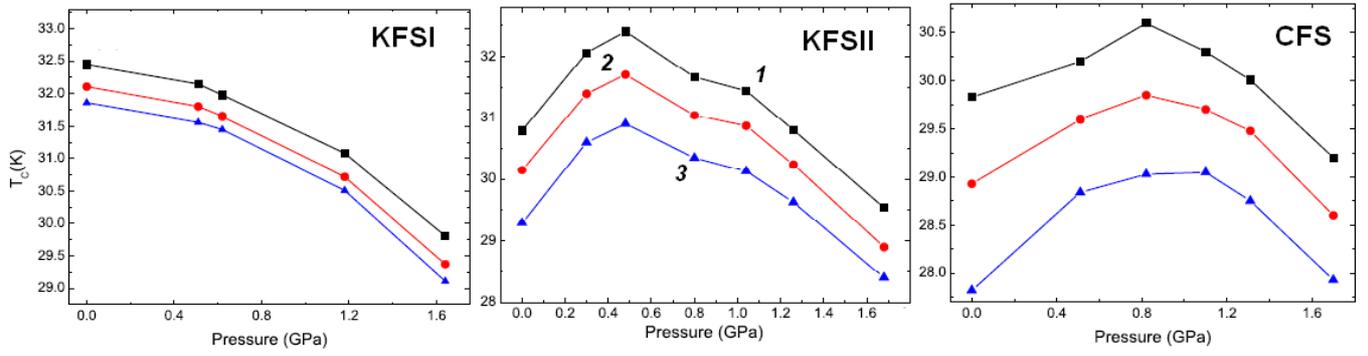

**Fig. 18**. Pressure dependence of transition temperature $T_C$ (at which resistivity drops by 90% (1), 50% (2), and 10% (3) relative to the resistivity immediately above the superconducting transition) for $K_{0.85}Fe_2Se_{1.80}$ (KFSI), $K_{0.86}Fe_2Se_{1.82}$ (KFSII), and $Cs_{0.86}Fe_{1.66}Se_2$ (CFS) [85].

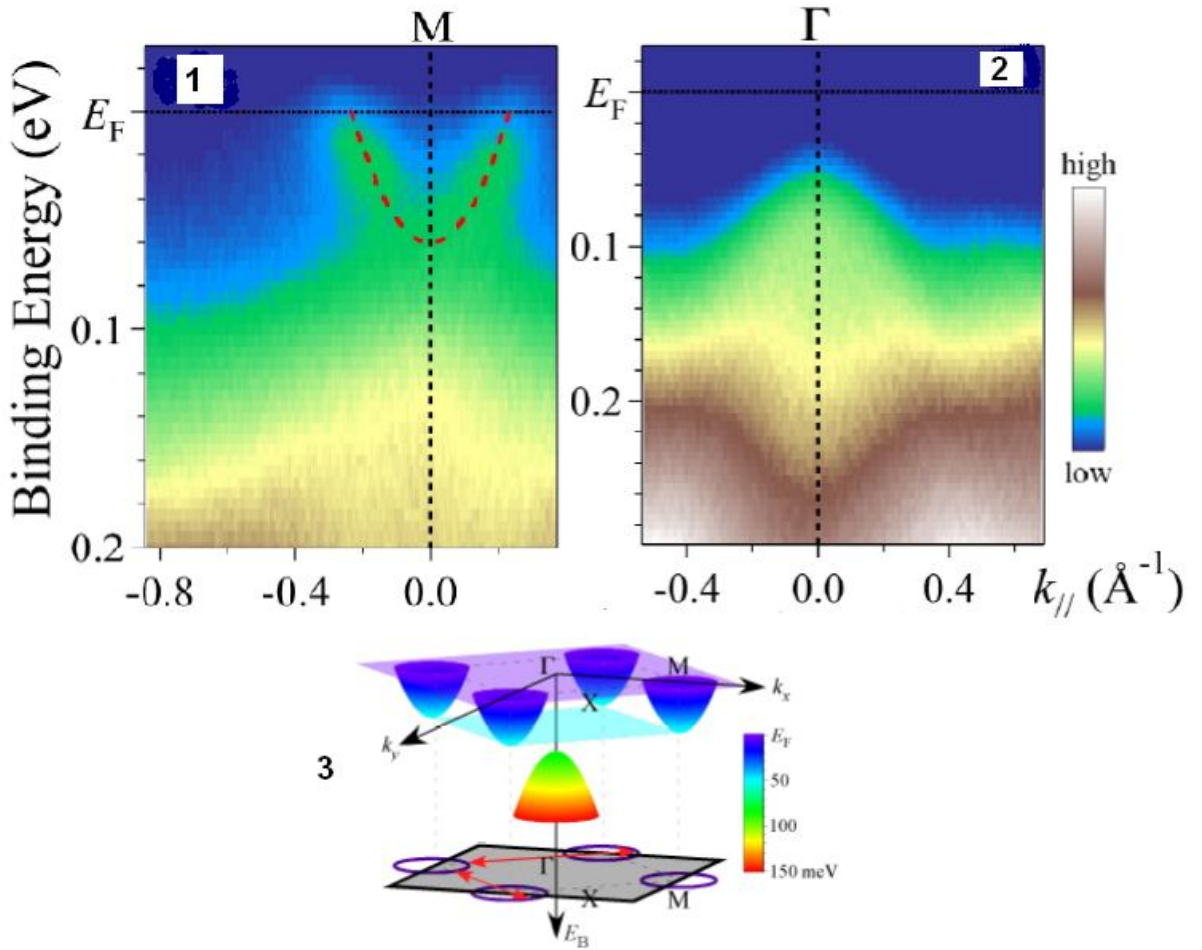

**Fig. 19.** ARPES intensity plots along a cut passing through M point (1) and along a cut passing through Γ point (2), and a schematic diagram summarizing the ARPES data on the electronic band structure of $K_{0.8}Fe_{1.7}Se_2$ and illustrating the (π, π) scattering process (3, *see also text*) [55].

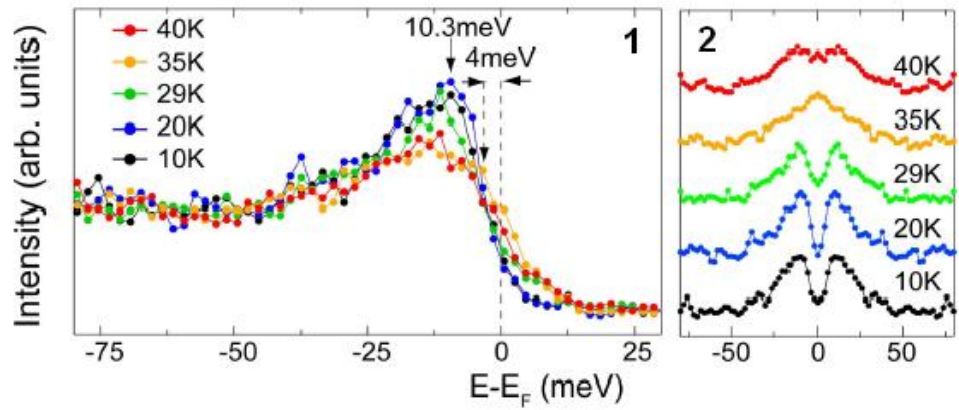

**Fig. 20.** Temperature dependence of the ARPES spectrum of $K_{0.8}Fe_2Se_2$ (1) and its symmetrized version (2) [54].

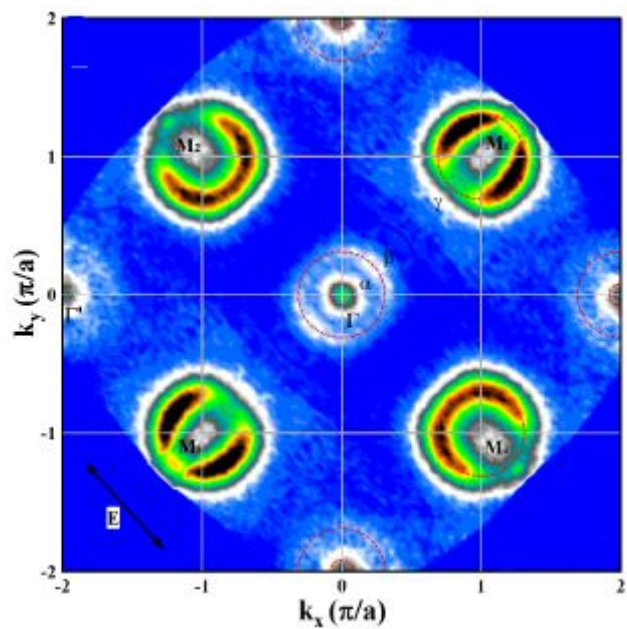

**Fig. 21**. Fermi surface of $Tl_{0.58}Rb_{0.42}Fe_{1.72}Se_2$ as obtained by ARPES [56].

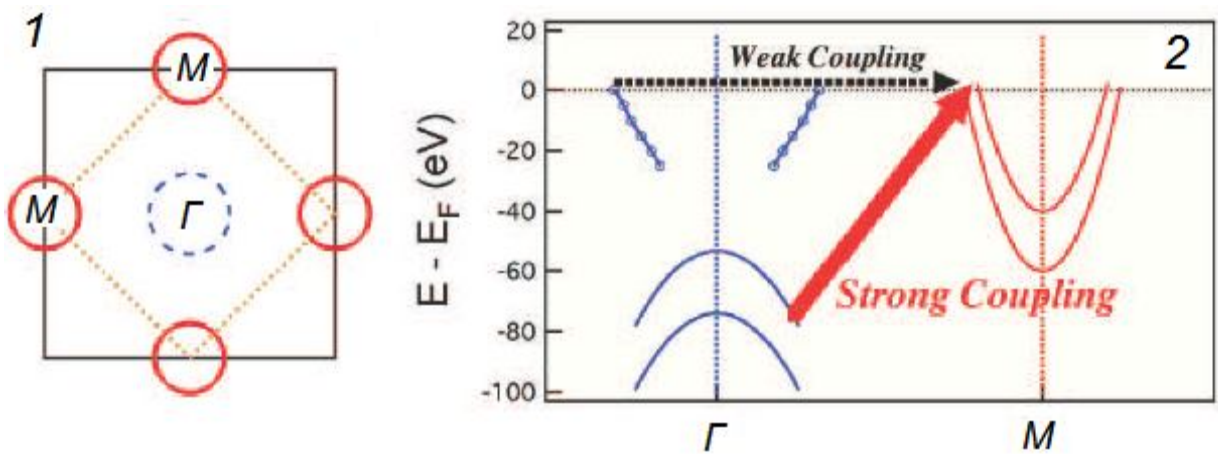

**Fig. 22.** Schematic: (1) Fermi surface and (2) band structure of $Tl_{0.63}K_{0.37}Fe_{1.78}Se_2$ as obtained within ARPES [57].

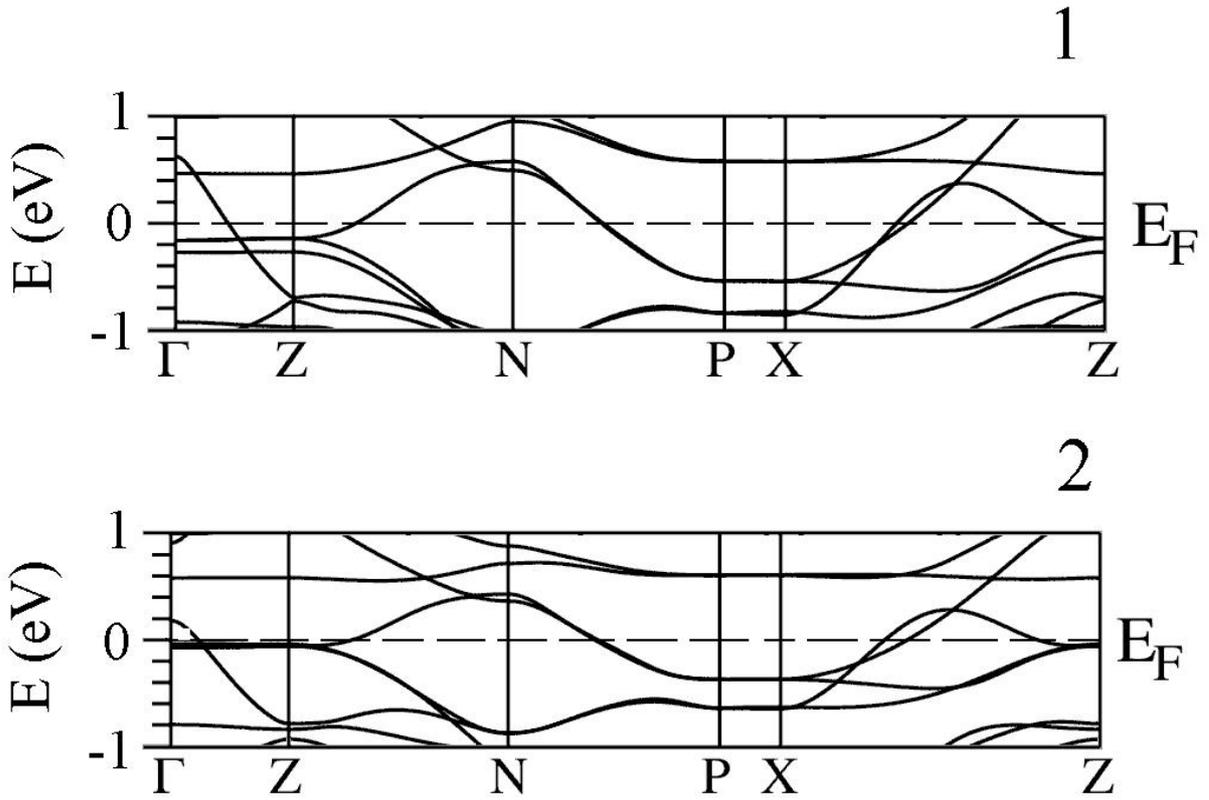

**Fig. 23.** Near-Fermi electronic bands for KFS$^{calc}$ (1) and KFS$^{exp}$ (2) [58].

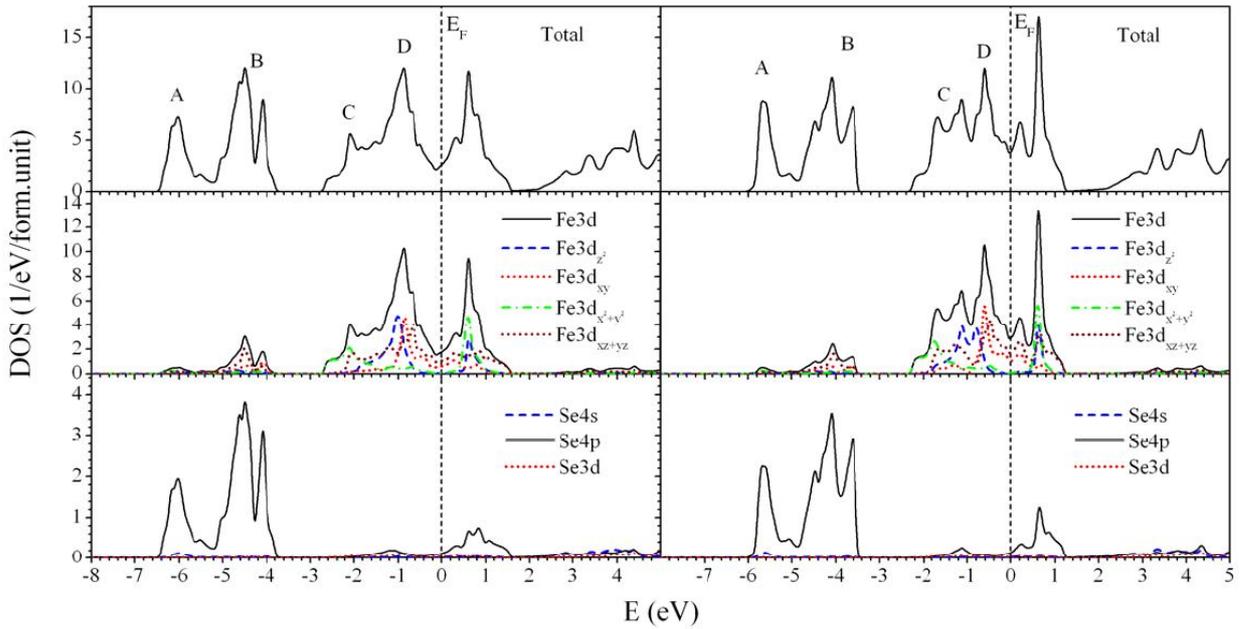

**Fig. 24.** Total (*upper panels*) and partial densities of states for KFS$^{calc}$ (*left*) and KFS$^{exp}$ (*right*) [58].

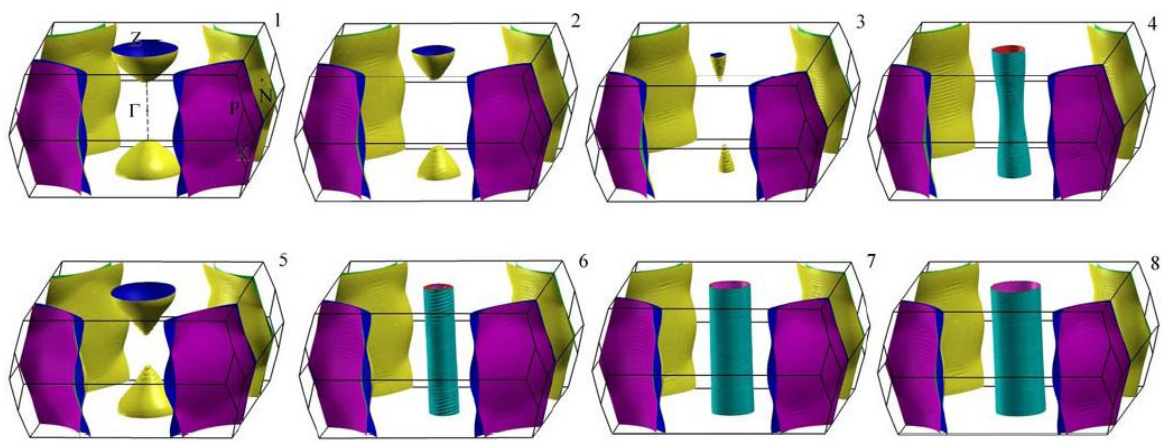

**Fig. 25.** Fermi surfaces for KFS$^{calc}$ for x = 1.0 (1), 0.8 (2), 0.7 (3), and 0.6 (4) (*top panel*) and for KFS$^{exp}$ for x = 1.0 (5), 0.8 (6), 0.7 (7), and 0.6 (8) (*bottom panel*) [58].

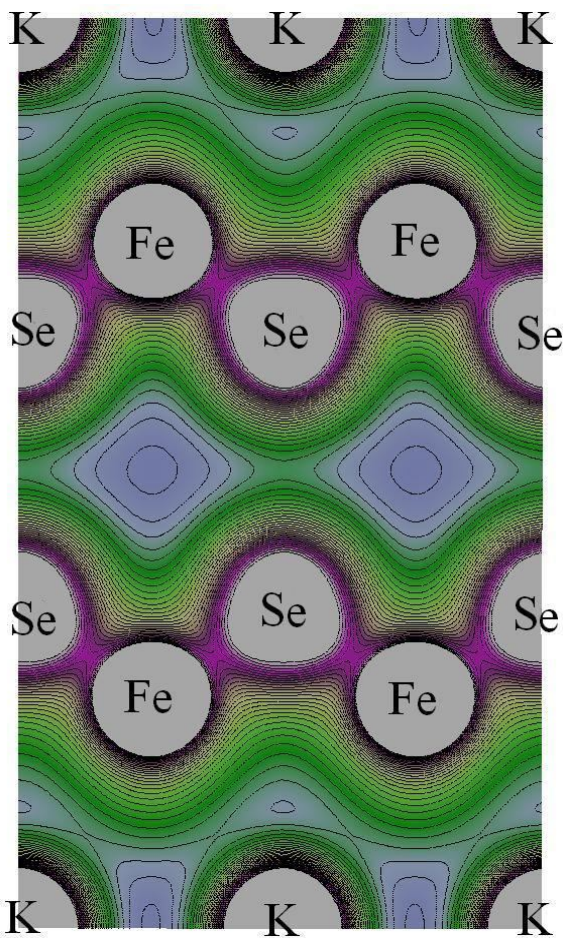

**Fig. 26.** Charge density map in the (001) plane for $KFe_2Se_2$ [93].

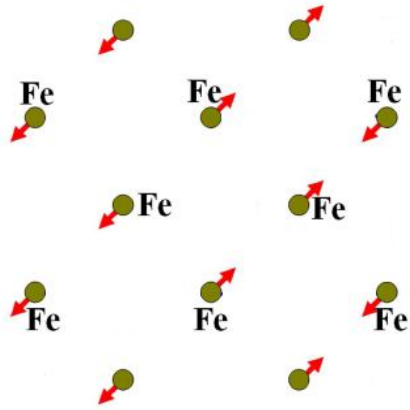

**Fig. 27.** The bi-collinear AFM ordering showing that the Fe moments align ferromagnetically along one diagonal direction and antiferromagnetically along the other diagonal direction in the Fe-Fe square lattice.

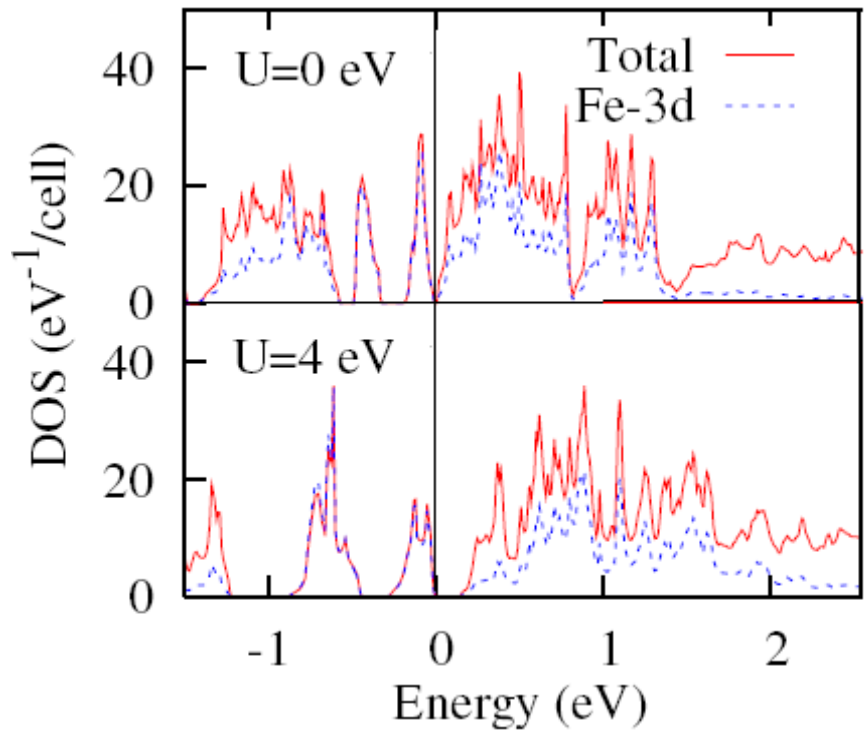

**Fig. 28.** Total and Fe-projected density of states of $TlFe_{2-x}Se_2$ as calculated for $U = 0$ eV and $U = 4$ eV [63]. The opening of the band gap with an increase in $U$ is obvious.

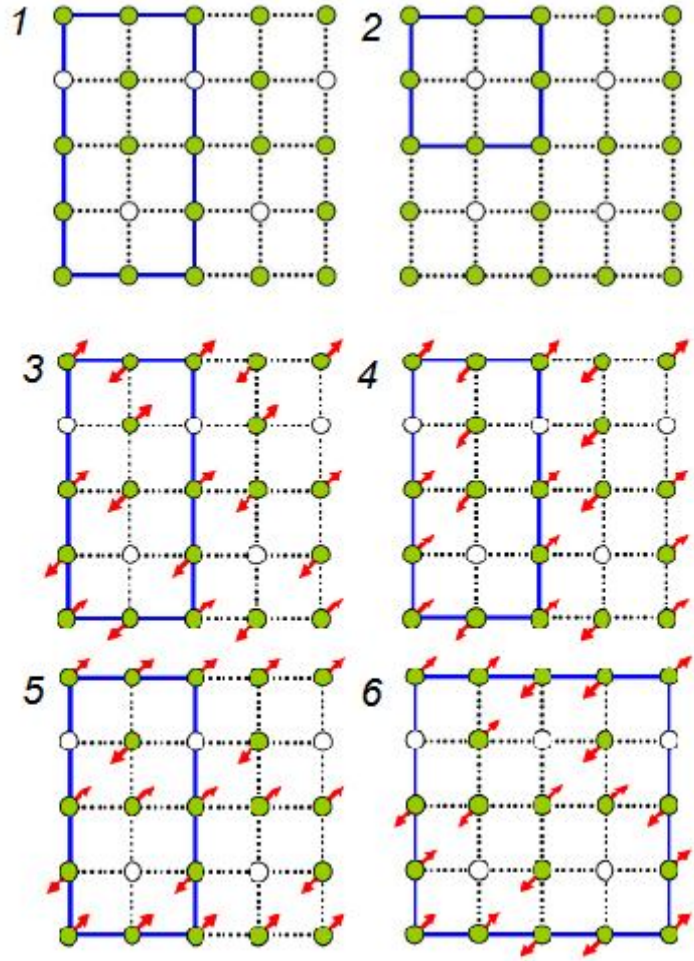

**Fig. 29.** *Top panel*: Schematic view of the Fe-Fe sheet with one quarter Fe-vacancies ordered in rhombus (*1*) and in square (*2*). The solid lines denote the corresponding cells. *Bottom panels*: Schematic view of four possible magnetic structures in the Fe-Fe sheet with one quarter Fe-vacancies ordered in rhombus: (*3*) Neel order, in which the nearest neighboring Fe moments are anti-parallel; (*4*) A-collinear AFM order, in which the Fe moments are ordered along the line without vacancies; (*5*) P-collinear AFM order, where Fe moments are ordered along the lines with vacancies; and (*6*) bi-collinear AFM order. The solid lines denote the magnetic cells [62].